\documentstyle[aps,pre,multicol,epsf,amssymb]{revtex}

\def\gs{\gtrsim}
\def\ls{\lesssim}
\def\gdot{\dot{\gamma}}
\def\be{\begin{equation}}
\def\en{\end{equation}}                  
\def\bea{\begin{eqnarray}}
\def\ena{\end{eqnarray}}
\newcommand{\bi}[1]{\mbox{\boldmath$#1$}}
\newcommand{\av}[1]{\langle{#1}\rangle}

\begin{document}
\draft
\bibliographystyle{prsty}
\title{Entanglements in Quiescent and Sheared Polymer Melts}
\author{Ryoichi Yamamoto and Akira Onuki}
\address{Department of Physics, Kyoto University, Kyoto 606-8502, Japan}
\date{\today}
\maketitle

\begin{abstract}
We visualize entanglements in polymer melts using molecular dynamics 
simulation. 
A bead  at  an entanglement interacts persistently for long times 
with the non-bonded beads 
(those excluding the adjacent ones in the same chain).
The interaction energy of each bead with the non-bonded beads is 
averaged over a time interval $\tau$  much longer than 
microscopic times but shorter than the onset time of  tube 
constraints $\tau_{\rm e}$. 
Entanglements can then be detected as hot spots consisting of 
several beads with relatively large values of the time-averaged 
interaction energy. 
We next apply a  shear flow with rate much faster than the entangle motion. 
With increasing strain the chains take zigzag shapes  
and a half of the hot spots become bent. 
The chains are first stretched as a network but, as the bends approach 
the chain ends, disentanglements subsequently occur, 
leading to  stress overshoot observed experimentally. 
\end{abstract}

\begin{multicols}{2}
\narrowtext
\newpage 
\hspace{-1cm} 

\section{Introduction}

The dynamics of dense polymer melts 
has been a challenging subject 
in current polymer physics \cite{Ferry,degennes,doi,mcleish,rubinstein}.
While the near-equilibrium 
dynamics of short chains with $N<N_{\rm e}$ can be reasonably 
well described by the simple Rouse model, 
the dynamics of very long chains with $N>N_{\rm e}$ 
has not yet been well 
understood on the microscopic level 
since it is governed by  complicated entanglement effects. 
Here $N$ is the polymerization index 
or the bead number of a chain, 
and $N_{\rm e}$ is the average bead number  between consecutive 
entanglements. The reptation theory
\cite{degennes,doi,mcleish,rubinstein,Pearson,mead} is the most successful  
approach to date in describing the dynamics 
of entangled polymer chains in a surprisingly simple manner. 
It has been supported by  a number of experimental  
\cite{schleger,ebert}
and numerical
\cite{kremer,paul,smith,putz,gao,kroger,hess,kroger2003,aoyagi,Pad,Pad2} 
papers, 
where   experimentally accessible quantities such as 
the stress relaxation function $G(t)$, 
the incoherent dynamic scattering function,
the mean square displacement, and the viscosity   have been 
compared with the theoretical predictions \cite{doi}.
However,  $N_{\rm e}$ 
 estimated from   the mean square displacement 
and that from the plateau modulus were around  30 and 70, 
respectively \cite{putz},  where the difference of these 
two estimations indicates  inaccuracy of  the prefactors 
in the predicted formulae \cite{doi,larson}. 
These  simulations were not  direct 
 observations of entanglements, 
but  rather confirmed the existence of the 
entanglement effects  indirectly using 
the formulae of the reptation theory. 
Measurements of the dynamic structure factor 
$S(k,t)$ by the 
neutron spin-echo method \cite{schleger} gave information of the tube 
diameter $\propto N_{\rm e}^{1/2}$ 
in the reptation theory and the resultant $N_{\rm e}
$ was consistent with the estimated value 
from the plateau modulus (both being of order 100).

Understanding macroscopic rheological 
 properties  of polymer melts in terms of 
 microscopic molecular dynamics is 
 also of great importance,  where the mechanism of 
 shear-thinning  behavior is very different 
 depending on whether  $N<N_{\rm e}$ 
 or  $N>N_{\rm e}$ and  whether 
 the glass transition 
is approached or not.   Though still inadequate, 
extensive simulations have 
been performed in this direction
\cite{kroger,hess,kroger2003,aoyagi,Cummings,yo1,yo2}.

The reptation theory \cite{degennes,doi} 
treats entanglements  as 
discrete objects severely constraining 
the chain motions only through "tubes". However,  
the exact nature of entanglements is not yet known 
and it  is highly nontrivial 
 how to  detect   them "directly" 
  in simulations.  Therefore, it is at present impossible 
to determine $N_{\rm e}$  "precisely" from  simulations. 
We also point out that the reptation theory does not 
provide the scaling functions of the physical quantities 
applicable even for not large 
 $N/N_{\rm e}$. 
The main  aim of this paper is hence 
 to give attempts  to identify and visualize 
entanglements    on the microscopic level.  Here we mention  related 
simulations.  Ben-Naim {\it et al.} \cite{ben-naim} 
 could visualize   individual entanglements using 
 the fact that incidental contacts of the particles 
and entanglement contacts behave differently 
because entanglement constraints are long-lived. 
Gao and Weiner introduced a time-averaged atomic 
mobility and found that
intrachain atoms of relatively 
low mobility tend to cluster in group
along the chain for $N=200$, which  suggests 
the existence of entanglements \cite{gao}.
In  recent Kr\"oger and Hess' simulation 
for  $10 \le N \le 400$ \cite{kroger2003},  
the zero-shear viscosity 
$\eta$ (obtained at extremely weak shear) 
changed over from the behavior 
$\eta \propto N$ to the 
behavior $\eta \propto N^a$ 
with $a$ in the range of  3 and 3.5   
around $N=N_{\rm c}\sim 100$.
This  crossover polymerization index $N_{\rm c}$ 
should be comparable to $N_{\rm e}$ but they did not 
set $N_{\rm c}/N_{\rm e}=1$, 
 probably because of the lack of  the 
theoretical scaling formula  
$\eta=Nf_{\rm vis}(N/N_{\rm e})$ 
describing the Rouse-to-reptation crossover as a function of 
$N/N_{\rm e}$. In a similar simulation, 
 Aoyagi and Doi \cite{aoyagi}  
calculated the steady state viscosity and 
normal stress differences to examine 
nonlinear rheology for $N=100,200,$ and  400.

In Table \ref{table1},  we 
 summarize the simulations 
of freely jointed 
bead-spring (Kremer-Grest type) chains. 
The estimated values of 
$N_{\rm e}$ are only crude ones. 
Except for  ours in this paper, 
they were  obtained indirectly  by fitting of 
numerical data to the predictions \cite{larson}.
At present it is still difficult to perform 
sufficiently large  simulations 
with $N\gg N_{\rm e}$.

The organization of this paper is as follows. 
In IIA our model system and our numerical method 
will be explained. 
In IIB, in the range  $N=10-250$, 
 we will examine the time-correlation function 
of the end-to-end vector $C(t)$, 
the stress relaxation  function $G(t)$,
the zero-shear viscosity obtained from 
$\eta = \int_0^\infty  dt G(t)$, 
and the diffusion constant $D$. These data 
will indicate $N_{\rm e} \sim 100$ with the aid of the 
reptation theory in agreement with 
the previous simulations.
The rheological crossover from the Newtonian 
to shear-thinning   behavior 
will also be examined for various $N$ as in the work 
by Aoyagi and Doi \cite{aoyagi}.   
We here stress that $G(t)$  
 exhibits multi-scale relaxations over many 
decades in chain systems 
and its numerical calculation has been rare 
because it requires 
very long simulations \cite{Pad,Pad2,yo1,yo2}.  
We will then 
 present   time-averaging methods   of 
detecting entanglements in quiescent 
states   in IIC and in rapidly  sheared states 
in IID. Our  "direct" 
observations of 
entanglements will again  give  $N_{\rm e}\sim 100$ and  
 enable us to  examine how the stress  
 overshoot and  chain stretching  occur in transient 
states \cite{Pearson,Graessley,inoue}.

\section{Numerical Section}

\subsection{Model}

We used the bead-spring model \cite{kremer} for 
our polymer melts 
composed of  $M$ chains with $N$  particles or 
beads in a cubic box 
with volume $V$. The total  particle 
number $NM$  was  
$1000$ for $N=10, 25$, and $100$, 
and was  increased to $2500$  for $N=250$. 
All the particle pairs  interact via a truncated  
Lennard-Jones potential defined by \cite{kremer} 
\be 
U_{\rm LJ}(r)= 4\epsilon [(\sigma/r)^{12}-(\sigma/r)^{6}]
 + \epsilon
\label{eq:1}
\en  
The right hand side  
is minimum at  $r=2^{1/6}\sigma$ 
and the potential is truncated (or zero) for larger $r$. 
Using   the  repulsive part of the 
Lennard-Jones potential only in this manner, 
 we may prevent spatial overlap of the particles \cite{kremer}.  
The number density was fixed at  
$n= NM/V =1/\sigma^3$, and 
 the temperature was 
 kept at $T=\epsilon/k_{\rm B}$.  At this temperature 
there was   no glass-like 
 enhancement of the structural relaxation time, but 
 at $T=0.2\epsilon/k_{\rm B}$ the present model with  $N=10$  
 became glassy in our previous 
simulation \cite{yo2}.  Note that 
our density value is higher  than the widely used 
value $n=0.85/\sigma^3$ in the previous 
simulations
\cite{kremer,putz,kroger,hess,kroger2003,aoyagi,ben-naim} 
(With increasing $n$  
the free volume for particles decreases 
and hence   $N_{\rm e}$ in our case should be 
somewhat  shorter than in the previous simulations).
The   consecutive beads on each chain  
are connected by an anharmonic spring potential of the form, 
\be 
U_{\rm FENE}(r)= -\frac{1}{2}k_{\rm c} R_0^2 \ln[1-(r/R_0)^2] 
\label{eq:2}
\en  
where $k_{\rm c}=30\epsilon/\sigma^2$ and $R_0=1.5\sigma$. 
The bonded pairs in the same chain  thus interact 
via the sum of the two potentials, 
$U_{\rm T}(r)= U_{\rm LJ}(r)+U_{\rm FENE}(r)$, 
which has a deep 
 minimum  at $r=b_{\rm min}=0.96\sigma$. 
In our simulations    the actual bond lengths  
between two consecutive beads remained very close to 
$b_{\rm min}$ with deviations being 
at most  a few $\%$ of $b_{\rm min}$ 
even under rapid shearing. 
In fact, the expansion 
 $U_{\rm T}(r)-U_{\rm T}(b_{\rm min})\cong  478\epsilon 
 (r/b_{\rm min}-1)^2$ 
follows around the minimum, so the deviation of 
this potential from the minimum value 
becomes of order $\epsilon$  
even for $r-b_{\rm min} \sim 0.04\sigma$.
(This means that the thermal fluctuation of the bond lengths 
is of order $0.04 \sigma (k_{\rm B}T/\epsilon)^{1/2}$.) 
Hereafter we 
will measure space and time 
in units of $\sigma$ and $\tau_0=({m\sigma^{2}/\epsilon})^{1/2}$, 
respectively,  with 
$m$ being the particle mass, unless confusion may occur. 
We numerically solved  Newton's equations of motion
and took data after long equilibration periods 
of order $10^6$.

In quiescent cases, 
we imposed the micro-canonical condition with 
time step $\Delta t = 0.005$ under the periodic boundary condition. 
 In the presence of shear flow, we set $\Delta t=0.0025$ and kept   
the temperature at $\epsilon/k_{\rm B}$ 
 using the Gaussian constraint 
thermostat to eliminate viscous heating \cite{Allen,Evans}.
After a long equilibration period in a quiescent state 
in the time region $t < 0$, 
all the particles acquired a velocity $\gdot y$ 
in the $x$ direction at $t = 0$, 
and then the Lee-Edwards boundary 
condition \cite{Allen,Evans} 
maintained the simple shear flow. 
The periodic boundary condition was imposed in 
the $x$ direction. 
Steady sheared states were realized after 
transient behavior.

For the case $N=250$  the 
system length  $V^{1/3}$  is  $2500^{1/3} 
\cong 14$, which is of the same order as 
the end-to-end distance of the chains ($\simeq bN^{1/2}$). 
 For such a small system size  
 under  the periodic boundary condition, 
 however, it is not clear how accurately we can simulate 
  the dynamics of real entangled polymers.   
 In future work,  simulations are desirable in larger  systems 
where both $N \gg N_{\rm e}$  and 
  $V^{1/3} \gg  \sigma N^{1/2}$ are well satisfied.

\subsection{Crossover from 
Rouse  to Reptation Dynamics}

First we show  that our numerical results near equilibrium 
are consistent with the Rouse or reptation theory \cite{doi}. 
In Fig.1, for various $N$ on a semi-logarithmic scale,  
 we show the normalized time-correlation function 
of the end-to-end vector ${\bi P}= {\bi R}_N-{\bi R}_0$,   
\be 
C(t)= \av{{\bi P}(t+t_0)\cdot {\bi P}(t_0)}/
 \av{|{\bi P}(t_0)|^2} 
\label{eq:3}
\en  
which  is normalized such  that $C(0)=1$. 
 Here, because of the finite size of  our system,
  the denominator and numerator on  the right hand side 
  remain to be considerably dependent on  $t_0$ 
even after taking the averages 
over all the chains \cite{comment4}. 
Thus  they were also averaged 
  over the initial time $t_0$. 
The statistical 
(temperature-dependent) bond length $b$ is defined by 
\be 
\av{|{\bi P}(t_0)|^2}=(N-1)b^2 
\label{eq:4}
\en  
In the present case  $b$ is equal to  
$1.25 \sigma$ and is longer than the actual bond lengths 
$|{\bi b}_j|=|{\bi R}_{j+1}(t)-{\bi R}_{j}(t)| 
\cong b_{\rm min} (=0.96\sigma)$, where 
$b_{\rm min}$ was  introduced 
below eq.2 \cite{comment}.
For this time-correlation function  
both  the Rouse dynamics and the reptation dynamics 
predict the same simple functional form \cite{doi}. 
For $N\gg 1$ it is a scaling function of 
$t/{\tau_{\rm r}}$ in terms of a relaxation time 
$\tau_{\rm r}$ as 
\be 
C(t)   =  \sum_{{\rm odd}~p} \frac{8}{\pi^2p^2} 
 \exp (- p^2{t}/{\tau_{\rm r}})
\label{eq:5}
\end{equation}
where  the summation is over odd $p$ and $1\le p \le N-1$. 
Since the first term ($p=1$) in the summation 
is dominant for any $t$,@
this function decays nearly exponentially  
and  $\tau_{\rm r}$ may be determined from  
$C(\tau_{\rm r})\cong e^{-1}$. 
As shown in Fig.1, for $N=10, 25, 100,$ and $250$, 
we obtained 
\be 
\tau_{\rm r}=  270, 1850, 4.6\times 10^4, 
6\times 10^5
\label{eq:6}
\en  
or    $\tau_{\rm r}/N^2=  2.7, 3.0, 4.6, 
$ and $9.6$, respectively. 
The calculated curve for 
$N=100$ can be excellently fitted to the  theoretical 
function $C(t)$ in eq.5.  For the other values of $N$ 
the deviation  is 
at most of order $4\%$  around $t \sim 0.1\tau_{\rm r}$. 
However, at long times $t \gs \tau_{\rm r}$,   
good agreement between the calculated   and 
theoretical curves was  obtained for any $N$.

Theoretically \cite{doi},
 the relaxation  time $\tau_{\rm r}$ 
 should be equal to the Rouse relaxation time 
\be 
\tau_{\rm R} = \tau_{01} N^2 
\label{eq:7}
\en 
for $N\ll N_{\rm e}$ and 
to the reptation or disentanglement time \cite{doi}  
\be 
\tau_{\rm d} =  \tau_{01}'N^3/N_{\rm e}
\label{eq:8}
\en 
for $N\gg N_{\rm e}$.  Here 
$\tau_{01}= \zeta b^2/3\pi^2 k_{\rm B}T$ 
and $\tau_{01}'=\zeta b^2/\pi^2 k_{\rm B}T$  
in terms of the monomeric friction constant 
$\zeta$ and the bond length $b$ in eq.4. 
If $\zeta$ and $b$ are assumed to be 
independent of $N$, the theory predicts 
\be 
\tau_{01}'= 3 \tau_{01}, \quad 
\tau_{\rm d}/\tau_{\rm R}= 3 N/N_{\rm e} 
\label{eq:9}
\en  
It has been argued that 
the Rouse time $\tau_{\rm R}$ has a well-defined physical 
meaning  even in the reptation regime 
as the relaxation time of 
the chain contour in a tube \cite{doi,Pearson,mead}.

For $N=10$,   the Rouse dynamics 
should be valid and  $\tau_{\rm r}=\tau_{\rm R}$
should hold,  so  our  data  of $b$ and $\tau_{\rm r}$ 
yield  
\be 
\tau_{01} \cong   2.7, 
\quad 
\zeta b^2/k_{\rm B}T \cong 80
\label{eq:10}
\en  
For the other values of $N$ 
we  have $\tau_{\rm R}= 0.027N^2$. 
Our  value of $\zeta$ 
is about  twice larger  than 
in the previous simulations 
with $n\sigma^3=0.85$ (see Ref. \cite{putz}).  
We then discuss the case of $N=250$, 
for which  our data give 
$\tau_{\rm r}/\tau_{\rm R} \sim 
4.6\times 10^4/
(0.027\times 250^2) \cong 3.6$ 
and the theoretical results  in eq.9  give  
 $\tau_{\rm d}/\tau_{\rm R} = 750/N_{\rm e}$.
In this paper (from  Figs.2, 5-7, and 10 below), 
we will obtain  $N_{\rm e}\sim 100$, which then yields 
$\tau_{\rm d}/\tau_{\rm R} \sim 7.5$ 
and $\tau_{\rm d}/\tau_{\rm r} \sim 7.5/3.6 \sim 2$.
Thus there arises a  difference of a factor 2 
between  the theoretical 
$\tau_{\rm d}$ and the numerically obtained 
$\tau_{\rm r}$. It  
could  stem from  two possible origins. 
One is that  
the Rouse-to-reptation crossover 
has not yet been well realized for $N=250 \sim 2.5 
N_{\rm e}$.
The other is that 
the prefactors in the reptation theory 
are inaccurate as suggested by P\"utz {\it et al.} 
for large $N$ \cite{putz}.
 These   seem to be both relevant  
in  the present work.

In Fig.2, we show our numerical data 
of the stress relaxation function for $N=25$ and 250 
expressed by 
\be
G(t) = (k_{\rm B}T)^{-1} V\langle \sigma_{xy}(t+t_0) 
\sigma_{xy}(t_0)\rangle
\label{eq:11}
\en 
where $V$ is the system volume   and 
the tensor  $\sigma_{\alpha \beta}(t)$ is defined by \cite{kroger,hess,yo2}
\be 
V^{-1} \int d{\bi r} \Pi_{\alpha \beta}({\bi r},t) 
= p(t) \delta_{\alpha \beta} - \sigma_{\alpha \beta}(t)  
\label{eq:12}
\en 
in terms of   the  microscopic  
 stress tensor $\Pi_{\alpha \beta}({\bi r},t)$. 
 The pressure $p(t)$ may be defined such that 
$\sigma_{\alpha \beta}(t)$ is traceless or  deviatoric.   
See Refs. \cite{kroger,hess,yo2} for the microscopic 
expression of $\sigma_{\alpha \beta}(t)$. 
On one hand, the curve for the short chain case $N=25$ 
and $M=40$ is a result of  
the averages over 10 independent runs and over 
the initial time $t_0$.  For $t\gs 10$  it  can 
be excellently fitted to the Rouse relaxation 
function (left dotted line) \cite{doi}, 
\be 
G_{\rm R}(t)=\frac{n k_{\rm B} T}{N} 
\sum_{p=1}^{N-1}
\exp(-p^2 t/\tau_{\rm 01}N^2)
\label{eq:13}
\en  
where $\tau_{\rm 01}=2.7$ as  determined in eq.10. 
However, for the longest  chain case $N=250$ and $M=10$, 
$G(t)$  was calculated  from a single very  long
run performed up to $ t= 5\times10^6=10^9\Delta t$ with 
the average over $t_0$ being taken. For $10 \ls t \ls 10^3$, 
it agrees with $G_{\rm R}(t)$ (right dotted line), but 
in the terminal time  range  
$t\gs \tau_{\rm r}= 6\times 10^5$ the calculated  $G(t)$ relaxes  
much slower than predicted by the Rouse dynamics. 
In Fig.2, although the characteristic  plateau behavior 
for $N\gg N_{\rm e}$ 
is  not clearly seen,  we write the  theoretical stress 
relaxation function $G_{\rm rep} (t)$ given by \cite{doi}  
\be 
G_{\rm rep} (t)=\frac{3n k_{\rm B} T}{5 N_{\rm e}} C(t)   
\label{eq:14}
\en 
with $N_{\rm e} = 100$ (dashed line),  
where   $C(t)$ is defined in  eq.5.    
The agreement appears to be fair for $t \gs 10^5$, but due to the 
noisy behavior and  the absence of well-defined  plateau 
in  $G(t)$,  
we may  claim only that  $N_{\rm e}$ is 
in the range $70\ls N_{\rm e}\ls 150$ from the fitting. 
Note that  the value of $\tau_{\rm r}$ at $N=250$ 
used for  $C(t)$ in   eq.14 
 is  obtained from  Fig.1 or eq.6.

Further remarks regarding Fig.2 are as follows. (i) First, the 
initial value $G(0)= V\av{\sigma_{xy}^2}/k_BT$  takes a very large value 
about  92 (not seen in Fig.2) and is nearly independent of $N$. 
Because  the initial values 
of $G_{\rm R}(t)$ and $G_{\rm rep}(t)$ 
are much smaller as $G_{\rm R}(0) \cong nk_{\rm B}T \sim 1$ and 
$G_{\rm rep}(0) \cong   3nk_{\rm B}T/5N_{\rm e} \sim 0.25$, 
agreement of $G(t)$ with these model relaxation functions is attained 
only after transient relaxations \cite{Onukibook}.
As reported previously \cite{Pad,yo1},
the high-frequency  vibration   
of the bond lengths (with period  $0.14$ here) gives rise to 
initial  oscillatory  behavior in $G(t)$ at  
short times ($t \ls 1$) \cite{comment2}.
(ii) Secondly, we notice  marked noisy behavior of 
the  curves in Fig.2  for $t \gs t_{\rm r}$ as already 
reported in Refs. \cite{Pad,yo2}. 
For  such large time separation,   the correlation 
of order $10^{-4}$ of the initial value 
needs  to be   picked  up, while 
 the amplitude of the 
thermal fluctuations of $\sigma_{xy}(t)$ 
at each time $t$ is  given by   
\be 
\sigma_{\rm fl}= \bigg [\av{\sigma_{xy}(t)^2}
\bigg ]^{1/2} \sim 
[G(0)k_BT /V]^{1/2}  
\label{eq:15}
\en 
This quantity is of order  $(92/2500)^{1/2}\sim 0.2$ 
for the case of $N=250$ in the dimensionless units \cite{yo2}.  
The noisy behavior in $G(t)$ in the terminal time range    
can in principle be eliminated only if  
the system  size is very large 
and/or many runs are performed.

In Fig.3, we summarize our results of 
 the relaxation time $\tau_{\rm r}$, 
 the zero-frequency shear viscosity 
$\eta = \int_0^\infty dt~ G(t)$ in the linear regime, 
and the diffusion constant $D$, as functions of $N$. 
Here  $\eta \propto N$ in the Rouse dynamics 
and $\eta \propto N^3/N_{\rm e}^2$ in the reptation  dynamics. 
The diffusion constant was  obtained 
from the relation, 
\be 
\av{|\Delta {\bi R}_{\rm G}(t)|^2} =6Dt  
\label{eq:16}
\en 
where $t>\tau_{\rm r}$  and  
 $\Delta {\bi R}_{\rm G}(t)=
 {\bi R}_{\rm G}(t_0+t)-{\bi R}_{\rm G}(t_0)$ is  the displacement  
 vector of  the center  of mass of the chains 
 in a time interval $[t_0, t+t_0]$ with width $t$. 
It is known that $D \propto N^{-1}$  in the Rouse dynamics 
and $D \propto N_{\rm e}/N^2$ in the reptation  dynamics \cite{doi}.
In Fig.3 we can see  that the dynamics of our system cannot be 
described by the Rouse dynamics with increasing $N$($\gs 100$). 
Though the largest $N$($=250$) is only 2.5 times larger than 
our estimated $N_{\rm e}$, the $N$ dependencies 
in Fig.3 are consistent with the  crossover from 
 the Rouse to reptation dynamics reported 
 in the previous simulations
\cite{kremer,paul,smith,putz,kroger,hess,kroger2003}.

We also show that 
our  model polymer melts exhibit strongly  
nonlinear  behavior for 
$\gdot > \tau_{\rm r}^{-1}$. 
Fig.4 displays  the steady-state 
shear viscosity $\eta$ 
in (a) and   the steady-state  normal-stress  difference 
$N_1$ in (b), where 
\be 
\eta (\gdot)= \av{\sigma_{xy}}/\gdot, \quad 
N_1(\gdot) = \av{\sigma_{xx}}-\av{\sigma_{yy}} 
\label{eq:17}
\en
Here the stress components consist of the contributions 
from all the particles as defined by eq.12 
and we  took 
the time-average 
$\av{\sigma_{ij}(t_0)}$ over  $t_0$ within 
an interval with width  $200$. 
The arrows 
indicate the points at which  $\gdot= \tau_{\rm r}^{-1}$. 
The $\eta(\gdot)$ at high  shear is 
 nearly independent of $N$ 
 in agreement with rheology experiments \cite{Ferry,mead,Stratton}  
and  the previous simulations \cite{kroger,hess,kroger2003,aoyagi},
while $N_1(\gdot)$ is an increasing function of $N$ at any 
shear rate.  If the data of $N=100$ and $250$ 
  at high shear are fitted to power laws, we roughly obtain 
   $\eta(\gdot) \sim \gdot^{-a}$  
and  $N_1 (\gdot) \sim \gdot^{a_1}N^{c_1}$  
with $a \sim 0.7$,  $a_1 \sim 0.5$, 
and $c_1 \sim 1.0$.  
  The second normal stress difference 
 $N_2(\gdot)= \av{\sigma_{yy}}- \av{\sigma_{zz}}$ 
(not shown here) 
 was  smaller than  $N_1(\gdot)$ by about 
 two orders of magnitude at any shear rates. 
In the literature \cite{doi,Pearson,mead},  it has been 
argued  that  the relaxation of 
the chain contours  occurs   on 
the time scale  of  $\tau_{\rm R}$ and   
is  of crucial  importance in nonlinear rheology 
under rapid deformations. 
In our case, however, Fig.1 shows that the ratio 
$\tau_{\rm r}/\tau_{\rm R}$ is only about 3.6   
 even for  $N=250$,   so we cannot draw definite 
 conclusions on the overall 
 nonlinear rheology.

\subsection{Entanglements in Quiescent States}

For  the longest chain case 
$N=250$, we  attempt  to identify and visualize 
entanglements. We  expect that there should be 
singular enhancement 
 in the  Lennard-Jones potential energy  between 
particles near  an  entanglement point. 
 To examine this effect, we first define the 
 potential energy of  non-bonded interaction 
on the $i$-th particle by 
\be 
W_i(t)= \sum_{j \in {\rm non-bond}}
U_{LJ} (|{\bi r}_i(t)-{\bi r}_j(t)|)  
\label{eq:18}  
\en 
The particle pairs $i$ and $j$  mostly  belong 
to different chains  giving rise to the inter-chain 
contributions  in eq.15. However, we also include 
the contributions from the pairs  belonging   to the same chain 
but being  not adjacent to each other ($j \neq i\pm 1)$.  
We found that the thermal fluctuations of 
$W_i(t)$ are so large 
that their distributions are 
nearly Gaussian at each time $t$ 
without any noticeable  correlations 
 even between adjacent beads ($i$ and $i+1$ on the same chain).
The non-bonded interactions 
 $W_i(t)$ consist mostly of 
rapidly varying thermal fluctuations uncorrelated 
to one another.  With the end of reducing  such  rapid components, 
we  introduce   the time average of 
 $W_i (t)$, 
\be
{W}_i(t,\tau)=\frac{1}{\tau} 
\int_0^{\tau} dt'  {W}_i (t+t')
\label{eq:19}
\en 
where the time interval $\tau$ is 
 much longer than the microscopic time 
$\tau_0$($=1$ in our units). This time average is analogous to that 
introduced by Thirumalai and Mountain in analyzing dynamics of  
supercooled liquids \cite{Mountain}.
We introduce  the  variance 
$\sigma (\tau)$ \cite{Mountain},
\bea 
\sigma(\tau)^2 &=&  \frac{1}{MN}
\sum_i    [{W}_i(t_0,\tau)- \av{{W}} 
]^2 \nonumber\\  
&=& \frac{2}{\tau} \int _0^\tau dt' (1- t'/\tau)
 F^{\rm NB} (t')  
\label{eq:20}
\ena 
where 
\be 
F^{\rm NB} (t')  = \av{{W}_i (t_0+t'){W}_i (t_0)} 
- \av{W}^2 
\label{eq:21}
\en 
Here $\av{W}$ is  
  the thermal average of $W_i(t_0)$ over all the beads 
nearly independent of $i$ and $t_0$. 
In the second line of eq.20,  
the time-correlation function $F^{\rm NB} (t')$ 
 is assumed to be independent of the initial time $t_0$.  
We found that $\sigma(\tau)^2$ 
decays nearly as $A\tau^{-1}$ for  $\tau  \gg 1$. 
The coefficient $A$ should then be given 
by $A= 2\int_0^\infty dt' F^{\rm NB}(t')$. 
This   indicates that 
most of the contributions in the time integral in eq.18  
 behave as thermal white noise \cite{Mountain}.

From the reptation theory \cite{doi,schleger} 
 the distance of the thermal 
bead motions 
during a time interval of 
$\tau$ is estimated in the short-time range
 $\tau \ll \tau_{\rm e}$ as  
\be 
\ell(\tau)=  \sqrt{\av{|\Delta {\bi R}(\tau)|^2}} 
\sim  a(\tau/\tau_{\rm e})^{1/4}
\sim  \tau^{1/4}
\label{eq:22}
\en 
where $\Delta{\bi R}(\tau)= 
{\bi R}(t_0+\tau)-{\bi R}(t_0)$ 
is the displacement vector of a bead during a time 
interval of $\tau$, 
$a= b N_{\rm e}^{1/2}$ 
is the tube diameter,  
and \cite{comment6} 
\be 
\tau_{\rm e}  
\sim \tau_{01}N_{\rm e}^2 \sim 3\times 10^4
\label{eq:23}
\en  
is the onset time of the effect of tube 
constraints. The relation  in eq.22 is 
well satisfied in the range $10 <\tau<10^4$ 
for $N=250$ in our simulation. 
This power law has been confirmed 
in the previous simulations \cite{kremer,paul,putz}. 
To achieve  visualization of 
 entanglements in the following, we should  require    
$\ell <  a \sim N_{\rm e}^{1/2}$ 
and  hence $\tau < \tau_{\rm e} \sim N_{\rm e}^2$ 
in eq.19.

With the above  time-averaging procedure, 
the white noise should be mostly eliminated 
with increasing $\tau$  and, as a result, 
 long-lived 
 correlations due to  a small number of entanglements 
should become detectable in the range 
$1\ll \tau < \tau_{\rm e}$.  
In  order to demonstrate this, in Fig.5,
we display normalized instantaneous 
values $[W_i(t)-\av{W}]/
\sigma(0)$ in (a) and 
normalized time-averaged values 
$[W_i(t,\tau)-
\av{W}]/\sigma(\tau)$ 
for   $\tau= 5\times 10^3=
0.8\times 10^{-2}\tau_{\rm r} \sim 0.2 \tau_{\rm e}$
in (b) at an appropriate  
time $t$ after a long equilibration period.  
For this $\tau$, the distance $\ell(\tau)$ in eq.22 is given by 4.5. 
Here the 10 chains in our system at time $t$ 
are straightened horizontally 
for the visualization purpose
(Similar pictures of  a time-averaged atomic 
mobility were given by Gao and Weiner \cite{gao}). 
In (a) almost no correlation can be seen along the chains.
In (b), on the other hand, 
 several consecutive beads form "active spots"   
having relatively large  values of 
$W_i(t,\tau)$.  
Here a bead (being the  $i$ th one 
in a chain) is defined to be active   if   
 the average
\be 
{\bar W}_i (t,\tau)= \frac{1}{7} 
\sum_{j=i-3}^{i+3} 
W_j(t,\tau)
\label{eq:24}
\en
over  the  7 adjacent  values 
($j=i-3, i-2, \cdots, i+3$)  in the same chain 
is larger than $1.1\sigma(\tau)$. 
This   averaging "in space"  
furthermore  eliminates random, 
small-scale fluctuations consisting of a few 
beads and, as a result, the variance of 
the average ${\bar W}_i (t,\tau)$ becomes 
$0.65\sigma(\tau)$ for $\tau= 5\times 10^3$. 
Then we select $4\%$ of 
the total beads as  active ones. 
In Fig.5b  they  form the  active spots   
numbered  from 1 to 18,   consisting 
 of several  consecutive active beads. 
In Fig.5c we show ${\bar W}_i (t,\tau)$ 
for the three chains with the active spots 
from $4$ to $7$.  Here we are expecting 
that   these active spots should arise from 
 entanglements in most cases except for accidental 
 enhancement of the non-bonded interactions. 
 Fig.5b indicates the existence of  
two or three  entanglements  on each chain
leading to the estimation  $N_{\rm e} \cong 100$.

To examine the  correlations 
in $W_i(t,\tau)$ along the chain contour 
quantitatively, we define the 
intra-chain  correlation function of 
the non-bonded interaction
\be
E(n,\tau)=\frac{1}{\cal{N}(\tau) } 
\sum_{j-i=n} [ W_i(t,\tau)- \av{W}] 
[W_j(t,\tau) - \av{W}] 
\label{eq:25}
\en
where the two beads $i$ and $j$ are 
separated by $n$ on the 
same chain and the average over all the chains is 
taken. The normalization factor $\cal{N}(\tau)$ 
is defined such that $E(0,\tau)=1$; 
then, ${\cal{N}}(\tau) 
 = MN  \sigma(\tau)^2 \propto \tau$.  
In Fig.6, we show 
$E(n,\tau)$ in (a) and its Fourier transformation  
\be
P(k,\tau)=\sum_{n=0}^{N-1}E(n,\tau)\cos(k n) 
\label{eq:26}
\en
in (b) for $\tau=0, 50, 500, 5\times 10^3$, and 
$5\times 10^4$. The longest $\tau$ is of the same order 
as $\tau_{\rm e}$ in eq.23.  
The most conspicuous feature is that 
 $E(n,\tau)$  takes a  negative 
minimum around $n = 45$ 
and  positive maxima around $n = 80$ 
with increasing $\tau$. The average  displacement 
 $\ell(\tau)$ in eq.22 was calculated  to be  $4.5$ for 
$\tau= 5\times 10^3$ and $7.7$ for 
$\tau=5\times 10^4$. 
 The contributions 
giving rise to these  extreme   
grow   in time in the normalized correlation 
$E(n,\tau)$ or  equivalently decay slower than $\tau^{-1}$ 
 in  the unnormalized
 correlation $ E(n,\tau){\cal{N} }(\tau)$. 
Correspondingly, the Fourier transformation 
$P(k,\tau)$ has a peak  at 
$k  \cong  2\pi/90$ at large $\tau$.  This suggests  
$N_{\rm e} \cong  90$.   
In addition, if $\tau$ is much larger than $\tau_{\rm e}$, 
 no periodic structure 
 was  observed in $E(n,\tau)$ (not shown here).
This should be because 
 the entanglements are 
delocalized along the chains 
in long time intervals with $\tau \gg  \tau_{\rm e}$.
Furthermore, as can be seen  in the inset of Fig.6a, 
the correlations  between nearby  beads 
$E(n,\tau)$ with $1 \le n \ls 10$ 
 are  nearly zero for 
$\tau=0$ but increases with increasing $\tau$.  
We may determine the characteristic 
width $n_{\rm w}(\tau)$  by 
$E(n,\tau)>10^{-2}$  
for $n<n_{\rm w}(\tau)$.  If this definition is used, 
the calculated values of  $n_{\rm w}(\tau)$ and  
$\ell(\tau)$ in eq.22 nearly coincide (within a few 10$\%$).

\subsection{ Entanglements   
under Rapid Shearing and Stress Overshoot}

Next, we applied   a  shear flow with  rate 
$\gdot=10^{-3}\sim 600/\tau_{\rm r} 
\sim 170/\tau_{\rm R}$ 
to the system of $N=250$ and $M=10$.  We  used  the same 
initial values for  the particle positions and momenta 
as those which produced   the data shown in Fig.5b. This  
is convenient to examine how  
entanglements behave in the quiescent 
and sheared conditions starting at exactly 
the same conditions.  
Here the time scale of the flow-induced chain deformations 
$( \sim \gdot^{-1}$) is much shorter than 
$\tau_{\rm e}$ in eq.23.  
In Fig.7 the chain conformations are projected 
onto the $xz$ plane (perpendicular to the 
velocity-gradient direction) 
at $\gdot t=5$ in (a) and $\gdot t=10$ in (b). 
In (b) the shear stress takes a maximum as will be shown 
in Fig.9. Because the chains are rapidly elongated, 
they eventually take zigzag shapes bent  
presumably at entanglements.  
The non-bonded interactions in these 
zig-zag points  become increasingly amplified with 
increasing  strain.  This should be  because 
a considerable fraction of the stress is supported by 
 entanglements in strong deformations.   
As a result, active spots in 
$W_i(t,\tau)$ 
 can be detected even with  much smaller $\tau$ 
than in the quiescent case, so we set $\tau=500=0.5/\gdot$ 
at this shear rate. In our simulation 
the  noise effect in $W_i(t,\tau)$ 
is much more reduced for  $\tau \sim \gdot^{-1}$ 
than in the quiescent case.  
Roughly speaking,  a $2/3$ fraction  of 
 the  numbered active  spots 
without shear   in Fig.5b remain to be active spots 
under shear strain of 0.5,  and 
a $1/2$  fraction  of them  
 become bent under shear in Fig.7a and Fig.7b. 
As in the criterion in Fig.5, 
the definition of the active beads 
is   given by ${\bar W}_i(t,\tau)
> 1.1\sigma(\tau)$ 
(but with much smaller $\tau$) and the number of the active 
beads is $4\%$ of the total bed number. 
We assign the same numbers 
to  these  active spots if  
their contour distance  between 
the locations along the chain 
in the quiescent and sheared cases 
remains shorter  than 10. The bend regions 
marked by $+$ in Fig.7, however, do not correspond to 
the numbered hot spots in Fig.5.

We can also see that the number of the bends 
has not  decreased from Fig.7a to Fig.7b,  
 but several of them are approaching the  chain ends and 
will  disappear (not shown here). 
Notice that the shear stress is maximum  at 
the time of Fig.7b.  In the reptation 
theory it is assumed that 
entanglements can be released 
only when they reach a chain end. 
If our bends represent entanglements, 
the disentanglement process induced by shear flow 
is going to start in Fig.7b, then leading to a decrease 
of the shear stress.

For comparison,  in Fig.8  we  show 
snapshots of the chains for the shorter 
chain case of $N=25$ in a quiescent state in (a) and 
under shear $\gdot=10^{-2}$ in (b), 
where $\tau_{\rm r}=1850(\cong 
\tau_{\rm R})$ and there is no entanglement.  
The non-bonded interactions are written with 
 $\tau=50$, but we cannot see any particular 
meaning in its heterogeneities on the chains 
with and without shear. 
In strong shear flow  $\gdot \tau_{\rm R} \gg 1$,  
the chains are at most times  elongated 
along the flow but undergo 
random  tumbling motions 
with period much longer than $\gdot^{-1}$. 
In (b)  compactly shaped chains are 
in the course of tumbling \cite{yo2}.

In the case of entangled melts, 
the shear stress $\sigma_{xy}(t)$ 
and the normal stress $N_1(t)= 
\sigma_{xx}(t)-\sigma_{yy}(t)$ are known to  exhibit
overshoot behavior in  rapid 
shearing \cite{Pearson,mead,kroger,hess,aoyagi,inoue}.
In Fig.9a we show their  time evolution  after application of shear 
at $t=0$ with  $\gdot=10^{-3}$ in the run of Fig.7. 
 On one hand, 
 $\sigma_{xy}(t)$ takes a maximum at 
  $\gdot t = 10$, at which the disentanglement starts.  
  On the other hand, $N_1(t)$ takes a maximum  
   at $\gdot t = 20$ afterwards. The same tendency of 
   successive maxima of $\sigma_{xy}(t)$ and $N_1(t)$  
   was  predicted theoretically \cite{mead} 
and observed 
   experimentally  under rapid shearing \cite{Pearson,Graessley,inoue}.
  In Fig.9b   we  show the numerical result of 
  the orientational tensor $Q_{\alpha \beta}(t)$ defined by  
\be
Q_{\alpha\beta}(t) = \frac{1}{M}  \sum_{\rm chain}  \frac{1}{N-1}
\sum_{j=1}^{N-1} b_{\rm min}^{-2} 
b_{j\alpha}(t)b_{j\beta}(t)   
\label{eq:27}
\en  
where   $b_{\rm min}^{-1} {\bi b}_j(t)$ are 
the normalized bond vectors and $\sum_{\alpha} 
 Q_{\alpha\alpha} \cong 1$ since 
 $|{\bi b}_j| \cong b_{\rm min}\cong 0.96\sigma$ 
 as stated below eq.2.  For this shear rate ($\gdot=10^{-3}$),  
  $Q_{xx}(t)- Q_{yy}(t)$ 
remains considerably smaller than 1 and 
 the chain bonds are still  weakly oriented along 
the flow direction.  
The overshoot of $Q_{xx}(t)- Q_{yy}(t)$  indicates 
retraction of the chain contours (tubes) after 
onset of disentanglement. 
Comparing Fig.9a and Fig.9b, we  notice  the proportionality 
of  the two deviatoric components  of the two  tensors 
\be 
 \sigma_{xy}(t)
 = A_0  Q_{xy}(t),  \quad
 N_1(t)=A_0 [Q_{xx}(t)- Q_{yy}(t)]   
\label{eq:28}
\en 
where  $A_0 = 2.2$ (with the stress components being measured 
in units of $\epsilon\sigma^{-3}$) \cite{comment5}.
Note that   the deviatoric part of the stress of polymer 
melts  is believed to be 
nearly equal to that of the  entropic 
stress contribution ($\sim k_{\rm B}T n Q_{\alpha\beta}$) 
 far above the glass transition 
 temperature \cite{degennes,doi}.
Furthermore, the deviatoric part  of 
the dielectric tensor $\epsilon_{\alpha\beta}$ is proportional 
to that of  $Q_{\alpha\beta}$ provided that 
 the microscopic polarization tensor 
is uniaxial along the bond direction. 
Thus we obtain   the well-known stress-optical relations   
\be 
\epsilon_{xy}= 
C_0 \sigma_{xy}, \quad \epsilon_{xx}- \epsilon_{yy}= 
C_0 (\sigma_{xx}-  \sigma_{yy})
\label{eq:29}
\en 
where $C_0$ is  a polymer-dependent 
constant.   In Fig.9c we also show the following  
angle
\be 
\chi= \frac{1}{2}  \tan^{-1} (2\sigma_{xy}/N_1) 
\label{eq:30}
\en  
On the basis of the 
stress-optical law, 
this angle is measured as the extinction angle 
$\chi= (1/2) 
\tan^{-1}[2\epsilon_{xy}/(\epsilon_{xx}- \epsilon_{yy})]$ 
in birefringence experiments. 
In Fig.8c we can 
see that $\chi$ exhibits a small  undershoot around $\gdot t=30$  
after the peaks of $\sigma_{xy}(t)$ and $N_1(t)$. 
A similar retarded undershoot was  observed experimentally 
 but  has not  been  explained 
theoretically \cite{mead}.

As the final example, 
we examine the case of 
  much larger shear rate $\gdot=10^{-2}\sim 6000/\tau_{\rm r} 
\sim 1700/\tau_{\rm R}$.  
Fig.10  displays  the snapshots of the chain conformations 
for $\gdot t=5$ in (a) and for $\gdot t=12.5$ in (b), where 
the chain stretching is stronger than in Fig.7 and 
the numbers of entanglements remain unchanged. 
The time interval $\tau$ in eq.16 is set equal to 50. 
In Fig.11 we show 
 $\sigma_{xy}(t)$ and 
$N_1(t)$ in (a), both exhibiting a peak around $\gdot t 
\cong  12.5$,  and  $Q_{xy}(t)$ and 
$Q_{xx}(t)-Q_{yy}(t)$ in (b). 
The  stress overshoot 
is more enhanced than in the smaller shear case 
in Fig.9, and the stress components decrease rather 
abruptly with onset of disentanglement. However, 
$Q_{xx}(t)- Q_{yy}(t)$  
saturates to a value about 0.7 without 
exhibiting overshoot.  Here even  the bonds 
 themselves align  in the flow direction 
and the chain stretching 
 becomes nearly complete. Interestingly, 
this bond alignment 
is still maintained even after onset of disentanglement.
  As we remarked below eq.2,  
 bond elongation of order $3\%$ 
is anharmonic for the potentials 
in eq.1 and eq.2 and 
gives rise to a tensile force of order $\epsilon/\sigma$. In this simulation, 
such strong forces are exerted on most of  the bonds 
 and  the proportionality relation in eq.25 does not hold.

\section{Summary and Concluding Remarks}

We have presented attempts to detect and visualize 
the entanglements in a model polymer melt 
together with attempts of indirectly 
deriving $N_{\rm e}$ as in the previous 
simulations. All the methods 
have yielded  $N_{\rm e} \sim 100$. 
We admit  that 
the visualization in   the  quiescent case is 
not yet firmly  established   by themselves 
in view of the fact 
that the thermal noise  still affects 
the data  even after   averaging 
in space and time as in Fig.5c. 
However,  under rapid  shearing,  
a large fraction of  active spots with relatively large 
nonbonded interactions 
 become bent, evidently indicating 
the existence of obstacles for the chain motion.  
Remarkably, the  active spots in the quiescent and sheared 
 cases  in Fig.5 and Fig.7 
 coincide with a large probability ($\sim 2/3$). 
We claim that a large fraction of such obstacles 
arise from entanglements  
preexisting  even before 
application of shear.     However, a few  
bends and hot spots  in Fig.7  
are not detected  by the visualization method in  Fig.5.  
This suggests that  some 
hot spots  from our method  
  may not represent  entanglements.

As remarked already at the end of IIA, 
the system size in our simulations 
is still comparable to 
the end-to-end distance for $N=250$  
and our results  need to be further checked 
in future large-scale simulations with longer chains.
To eliminate the large thermal noise effect in the quiescent case, 
we should take data in the well-defined 
reptation regime under $N \gg N_{\rm e}$.

Performing very long simulations, 
we have also calculated the stress 
relaxation function $G(t)$, which exhibits the 
Rouse-to-reptation 
crossover with increasing the polymerization index $N$, 
and studied nonlinear rheology in transient and steady 
sheared states.  
In transient states under shear,  
 the stress overshoot sets in 
 as the bends approach the chain ends and 
disappear, as can be seen from Fig.7b and Fig.9. 
This is also one of our main results 
giving molecular information on the stress overshoot 
under rapid shearing.

In real long chain systems, 
the ratio $\tau_{\rm d}/\tau_{\rm R} 
\sim N/N_{\rm e}$ can be very large. Hence, in shear flow, 
there can be three characteristic shear regions \cite{mead} 
given by (i) $\gdot < \tau_{\rm d}^{-1}$, 
(ii) $\tau_{\rm d}^{-1} <  \gdot < \tau_{\rm R}^{-1}$, and 
(iii) $\gdot> \tau_{\rm R}^{-1}$. 
Nonlinear shear effects emerge 
in the regions (ii) and (iii), 
while  the linear response theory in terms of $G(t)$ in  eq.11 
is valid only in the region (i). 
In our study, the intermediate region (ii) 
is not  well-defined, but the calculated overshoot and undershoot 
relaxations in Fig.9 (in the region (iii)) 
resemble those in the experiments \cite{Pearson,Graessley,inoue}.

This work is supported by 
Grants in Aid for Scientific 
Research from the Ministry of Education, Science, Sports and Culture
of Japan.
Calculations have been performed at the Human Genome Center, 
Institute of Medical Science, University of Tokyo and 
the Supercomputer Center, Institute for Solid State Physics, 
University of Tokyo.

\end{multicols}

\narrowtext


\newpage

\widetext
\begin{table}
\caption{Summary of simulations 
 of  freely jointed chains.
The second column  shows  the  chain length  in  the
simulations. The third  and fourth  columns
show the system volume $V$ and the simulation time 
$t_{\rm max}$  in units of the chain dimension  $bN^{1/2}$
 and the  Rouse time $\tau_0N^2$. In the fifth to tenth 
columns the estimated values  of $N_{\rm e}$  
are given (if estimated). 
\label{table1}
}
\begin{tabular}{cccccccccc}
Refs. & $N$ & $V^{1/3}/bN^{1/2}$ & $t_{\rm max}/\tau_0N^2$ & $N_{\rm e}$\tablenotemark[1] & $N_{\rm e}$\tablenotemark[2] & $N_{\rm e}$\tablenotemark[3] &$N_{\rm e}$\tablenotemark[4] & $N_{\rm e}$\tablenotemark[5] & $N_{\rm e}$\tablenotemark[6]\\
\tableline
KG1990 \cite{kremer}& $200$ & 0.9 & $1.5$ & $120$ & $35$ & $34\sim50$ & $60$ & &\\
KLH1993 \cite{kroger}& $300$ & 1.5 & $11$ & & & & & 100 & \\
GW1995 \cite{gao}& $200$ & 0.9 & $0.3$ & & & & 
&  & \\
BGWB1996 \cite{ben-naim}& $350$ & 1.9 & $1.8$ & & & & & & \\
AD2000 \cite{aoyagi}& $400$ & 0.6 & $0.4$ & & & & & & \\
PKG2000 \cite{putz}& $350$ & 1.5 & $8.2$ & & & $ $ & & & \\
               & $700$ & 1.9 & $0.2$ & & $32\sim35$ & $ $ & & & \\
               & $10000$ & 0.7 & $0.0002$ & & $28\sim32$ & $ $ & $65\sim83$ & & \\
KH2000 \cite{kroger2003}& $400$ & 2.7 & $63$ & & & & & 100 & \\
Present study
& $250$ & 0.7 & $80$ & $100$ & & & $100$ & $100$ & $90$ 
\end{tabular}
\tablenotetext[1]{from  center-of-mass diffusion}\tablenotetext[2]{from  monomer diffusion}
\tablenotetext[3]{from  scattering function}
\tablenotetext[4]{from  stress relaxation function}
\tablenotetext[5]{from  zero-shear viscosity}
\tablenotetext[6]{from direct visualizations}
\end{table}

\narrowtext
\begin{figure}[b]
\centerline{\epsfxsize=2.8in\epsfbox{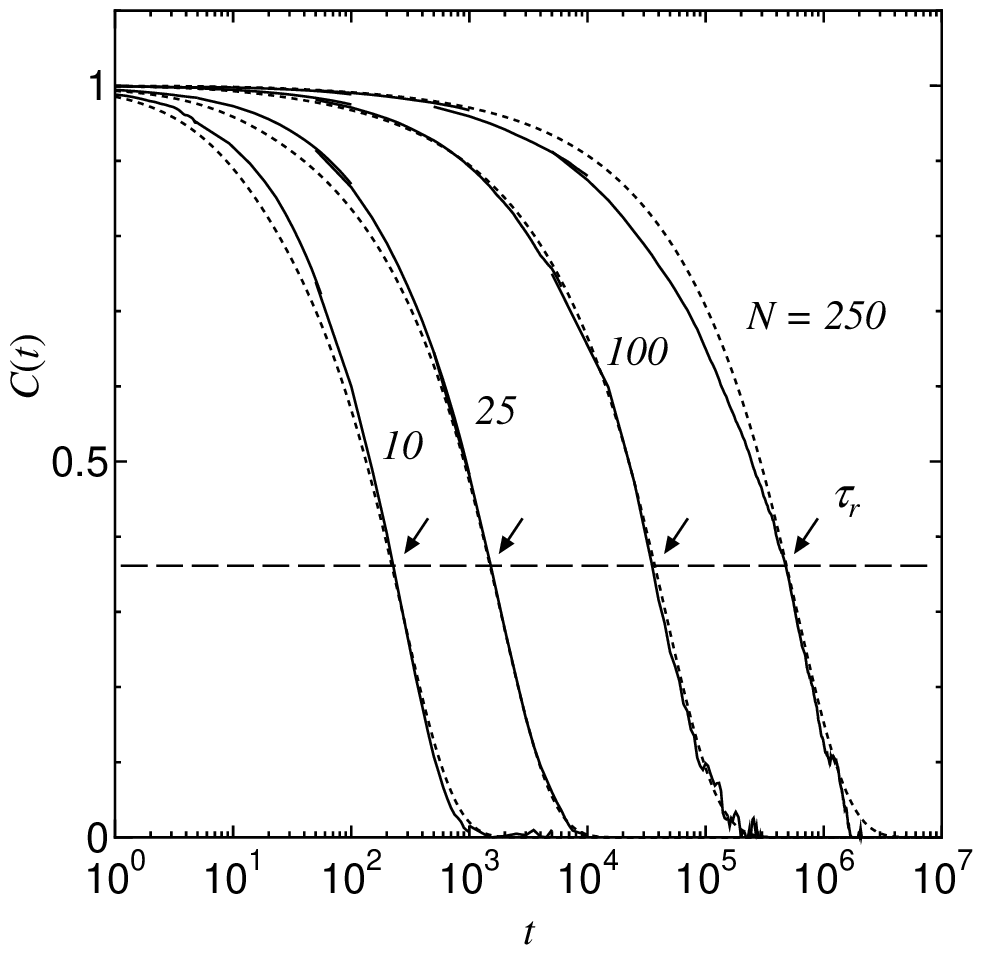}}
\caption{\protect
The normalized 
 time-correlation function 
$C(t)$ for the end-to-end vectors 
in eq.3  for various $N$ (solid lines) 
on a semi-logarithmic scale. They can be fitted 
to the theoretical expression in eq.5 (dashed lines). 
 The arrows indicate the terminal relaxation time 
 $\tau_{\rm r}$ defined by $C(\tau_{\rm r})=e^{-1}$, 
 which is the Rouse relaxation time $\tau_{\rm R}$ 
 or  the reptation time $\tau_{\rm d}$  depending 
 on whether $N\ll N_{\rm e}$ or $N\gg N_{\rm e}$.}
\label{fig1}
\end{figure}
 
\newpage
\narrowtext
\begin{figure}[b]
\centerline{\epsfxsize=2.8in\epsfbox{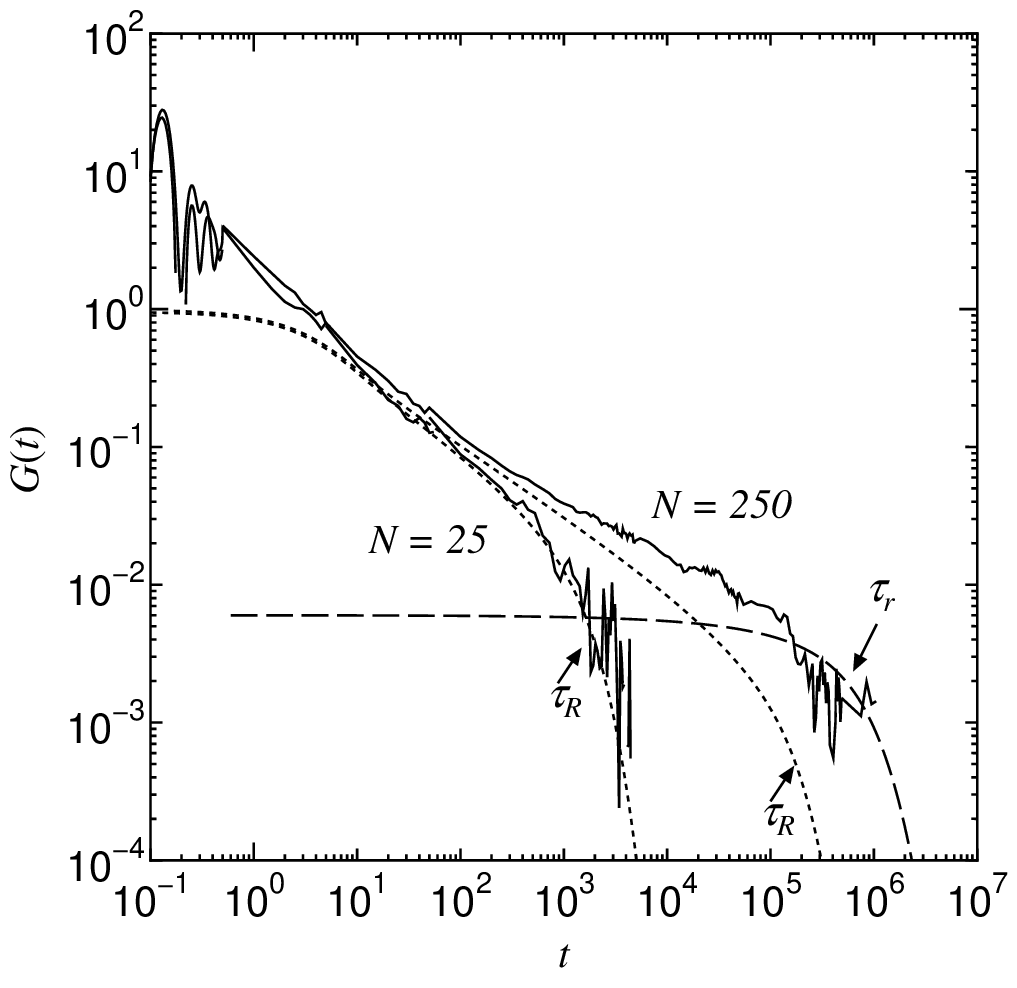}}
\caption{\protect
The stress relaxation function $G(t)$ 
 for $N=25$ and $250$ (solid lines) 
 in units of $\epsilon \sigma^{-3}$. Its 
  $N$ dependence  is weak for $t \ls 10^2$ 
 but becomes increasingly  stronger at later times.  
For $N=25$ the curve of $G(t)$ can well be 
fitted to the Rouse relaxation 
function $G_{\rm R}(t)$ 
with $\tau_{\rm 01}=2.7$ and $N=25$ in eq.10 (left dotted line). 
For $N=250$, it can be fitted to 
the reptation relaxation function   $G_{\rm rep}(t)$ 
in eq.12  for $t\gs \tau_{\rm r}$  (dashed line), where 
 $\tau_{\rm r}= 6\times 10^5$ in $C(t)$ 
and $N_{\rm e} = 100$ in the prefactor. 
For comparison 
we also show $G_{\rm R}(t)$ with 
$\tau_{\rm 01}=2.7$ and $N=250$ (right dotted line). It 
much deviates from the calculated curve of $G(t)$ 
of $N=250$ for  $t\gs 10^4$, 
while they agree fairly well for $10 \ls t\ls 10^3$.}
\label{fig2}
\end{figure}

\newpage
\narrowtext
\begin{figure}[t]
\centerline{\epsfxsize=2.8in\epsfbox{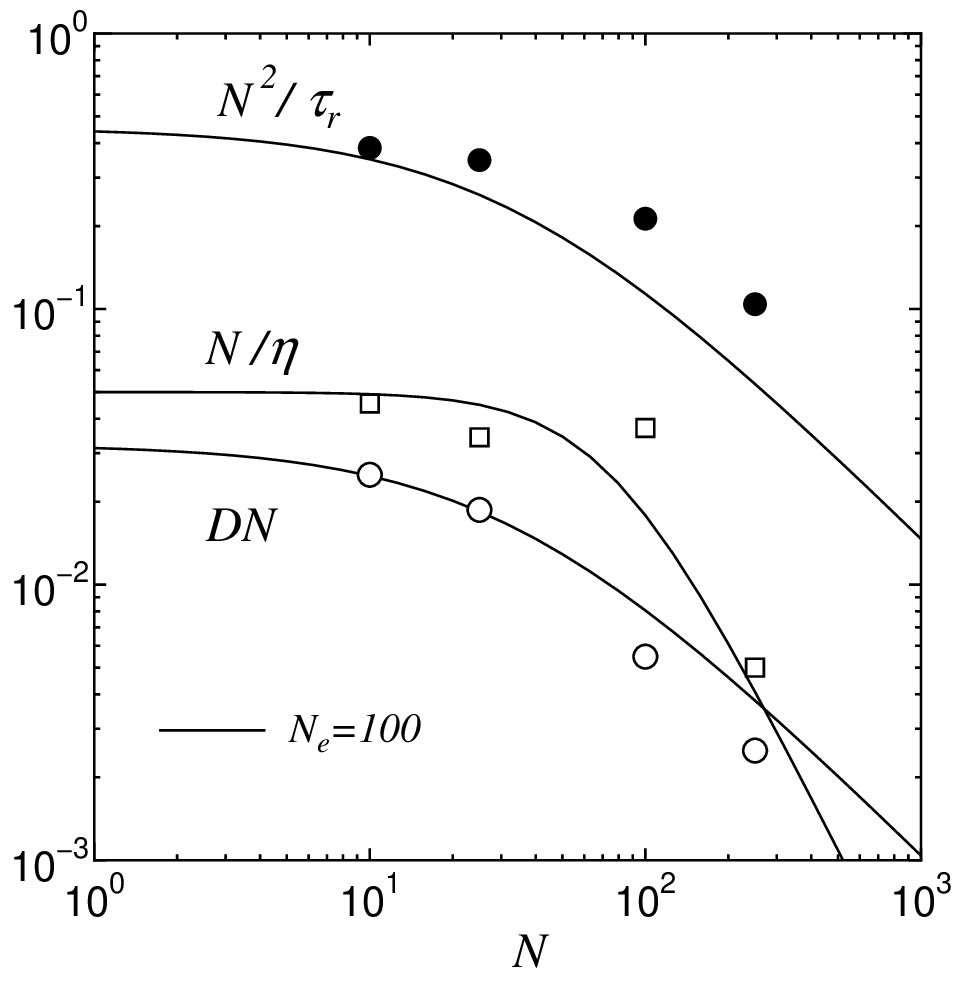}}
\caption{\protect
$N$-dependence of $N^2/\tau_{\rm r}$ ({\Large$\bullet$}), 
$N/\eta$ ({\small$\blacksquare$}), and $ND$ ({\Large$\circ$}). 
These quantities are expected to 
tend to the Rouse limits 
for $1 \ll N \ll N_{\rm e}$. 
The solid  lines represent  
$(3\pi^2k_BT/\zeta b^2)/(1+ 3N/N_{\rm e})$, 
$(36/n\zeta b^2)/(1+8N^2/5N_{\rm e}^2)$, and 
$(k_BT/\zeta)/(1+3N/N_{\rm e})$ with $N_{\rm e}=100$. 
These   approximate expressions extrapolate 
the predicted formulae$^3$ of the 
Rouse and reptation behaviors in the two limits 
$N \ll N_{\rm e}$ and $N \gg N_{\rm e}$. 
}
\label{fig3}
\end{figure}

\newpage
\widetext
\begin{figure}[t]
\centerline{\epsfxsize=2.8in\epsfbox{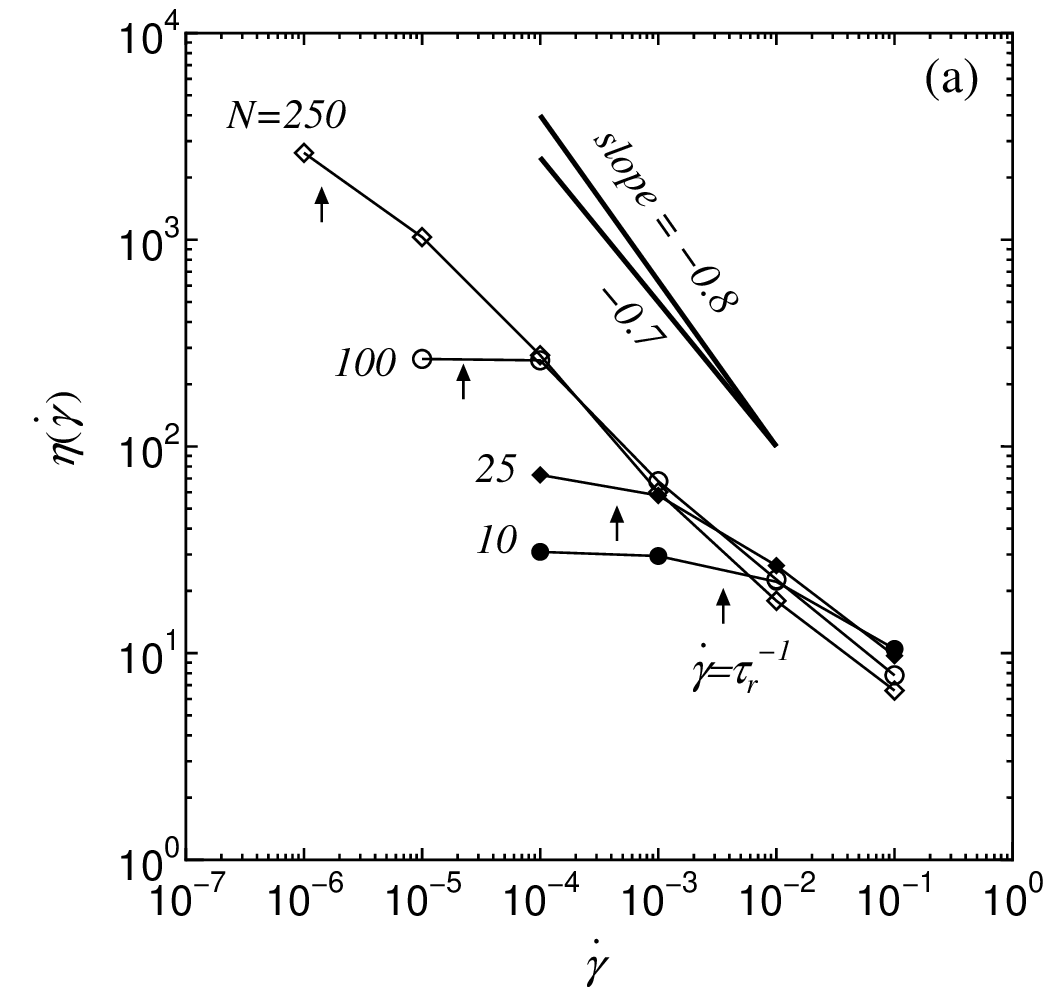}\epsfxsize=2.8in\epsfbox{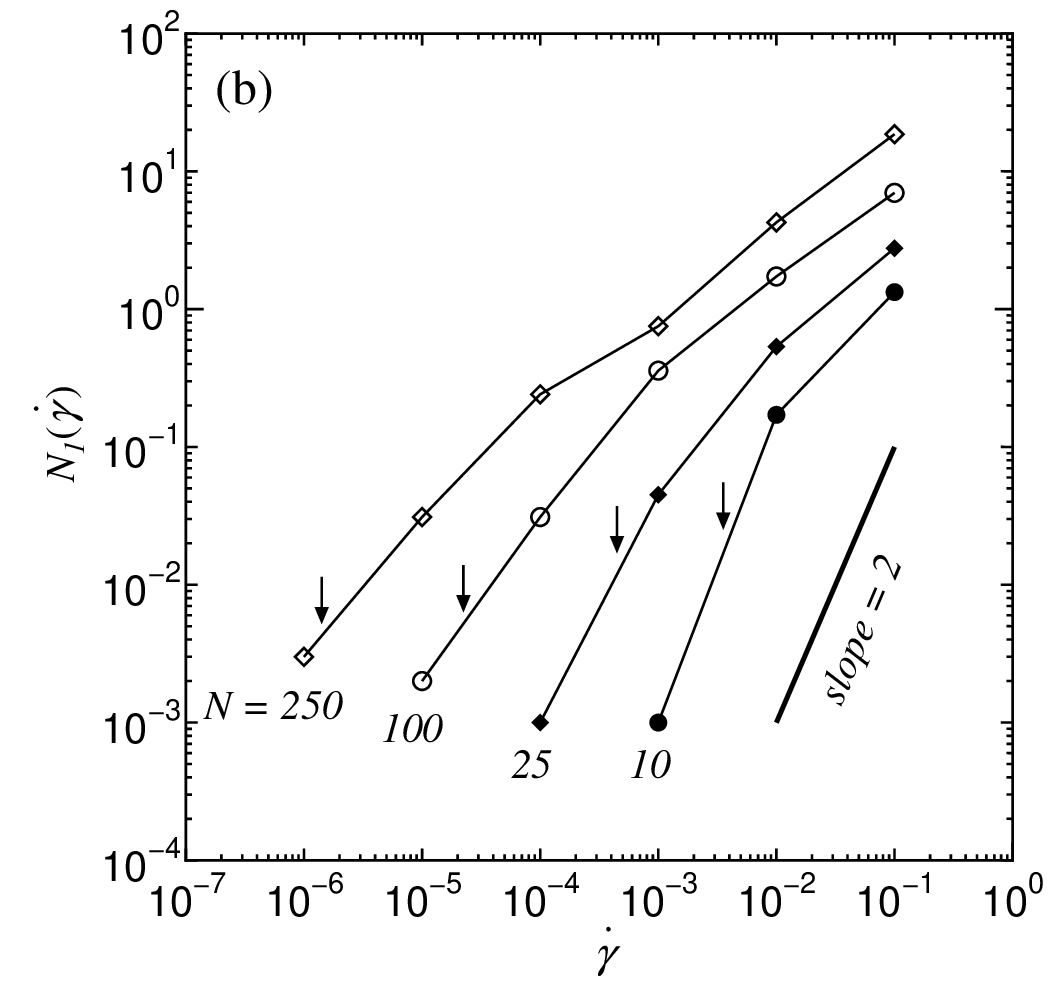}}
\caption{\protect
 Calculated 
 viscosity $\eta(\gdot)=\sigma_{xy}/\gdot$ in (a),  
 and normal stress difference $N_1(\gdot)=
 \sigma_{xx}-\sigma_{yy}$  in (b) 
 in steady states in shear for various $N$. 
The arrows indicate the points at which 
$\gdot =\tau_{\rm r}^{-1}$. As guides of 
 eyes,  we write lines of slope -0.7 and -0.8 in (a) 
 and a line of slope 2 in (b).  
 Note that $N_1(\gdot) 
\propto \gdot^2$ for Newtonian shear. }
\label{fig4}
\end{figure}

\newpage
\widetext
\begin{figure}[b]
\noindent
\centerline{\epsfxsize=3.5in\epsfbox{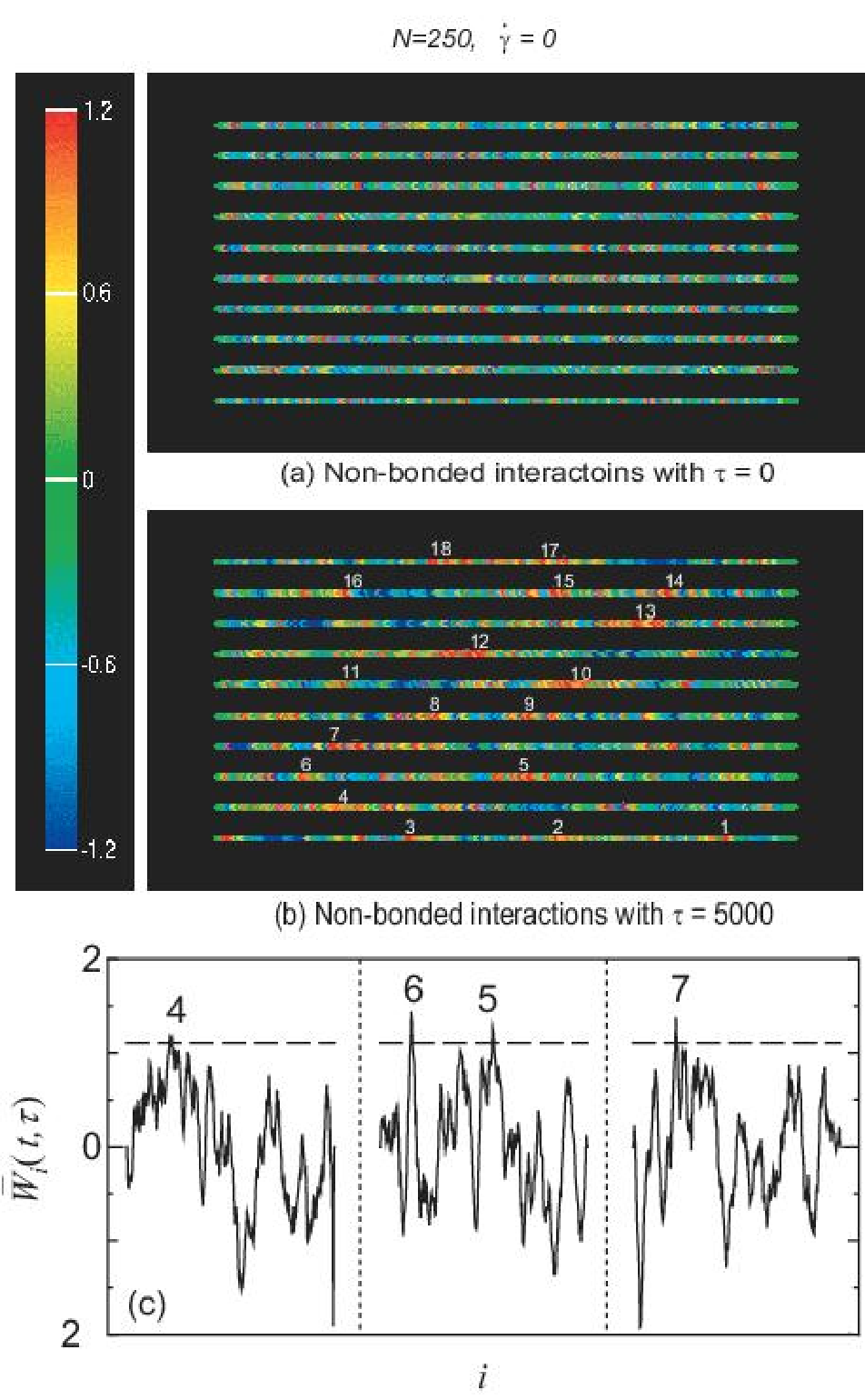}}
\caption{\protect
Distributions of 
the  non-bonded interaction energy  on the 
chains for $N=250$ in a quiescent state. 
The 10 chains are straightened on the plane.  
In (a) the normalized  values 
$[W_i(t)- \av{W}] /\sigma (0)$ 
  are shown, where $\tau=0$ and 
no correlations along the chains 
can be seen. In (b) 
 the  normalized, time-averaged   values 
 $[W_i(t,\tau)- \av{W}] /\sigma (\tau)$ with  $\tau = 
5 \times 10^3$  are shown, 
which are distinctly large  in line segments  
 consisting of several beads 
(in orange) presumably due to entanglements. 
In (b) the active spots are numbered from 
1 to 18 according to the criterion 
 given around eq.24.  In (c) the data of 
$[W_i(t,\tau)- \av{W}] /\sigma (\tau)$  
for the three chains with the spots 4-7 in (b) are shown. 
The horizontal axis denotes the bead numbers $1\le N \le 250$ 
for the three chains. The beads above the broken line 
are defined as "active" beads. }
\label{fig5}
\end{figure}

\newpage
\widetext
\begin{figure}[t]
\centerline{\epsfxsize=2.8in\epsfbox{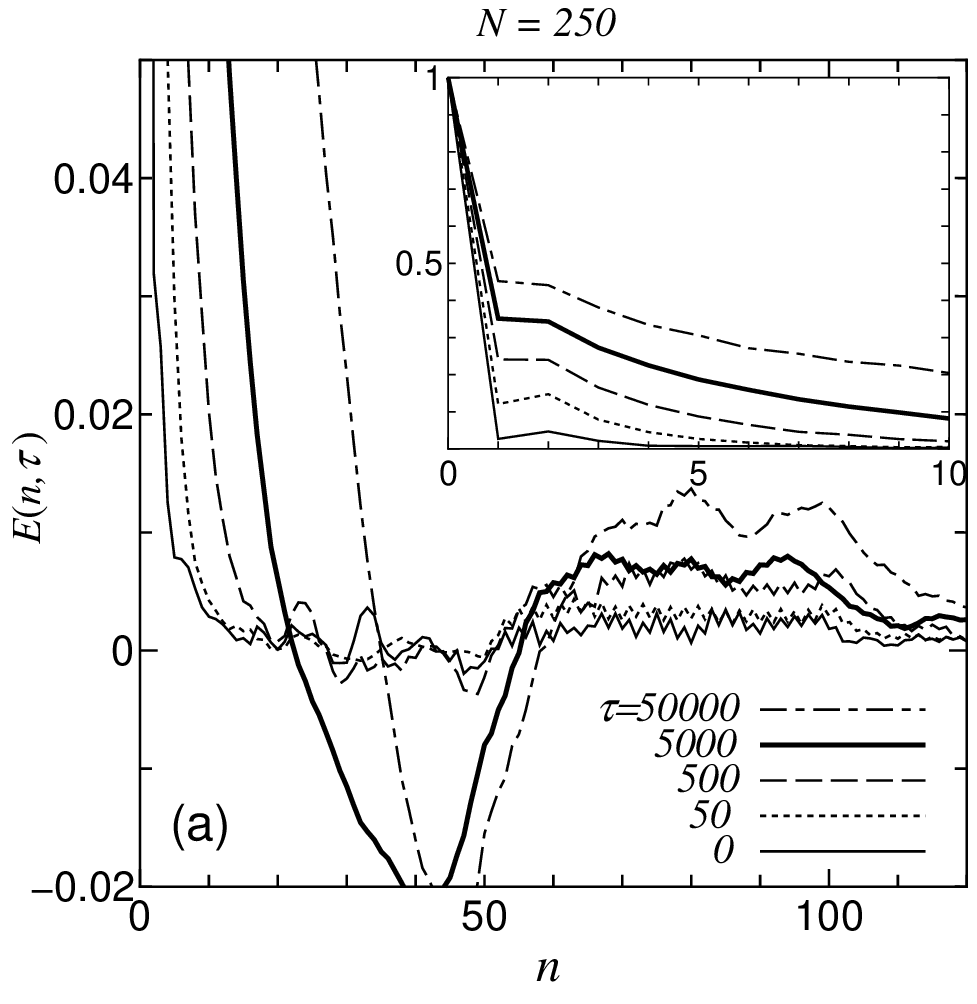}\hspace{10mm}\epsfxsize=2.8in\epsfbox{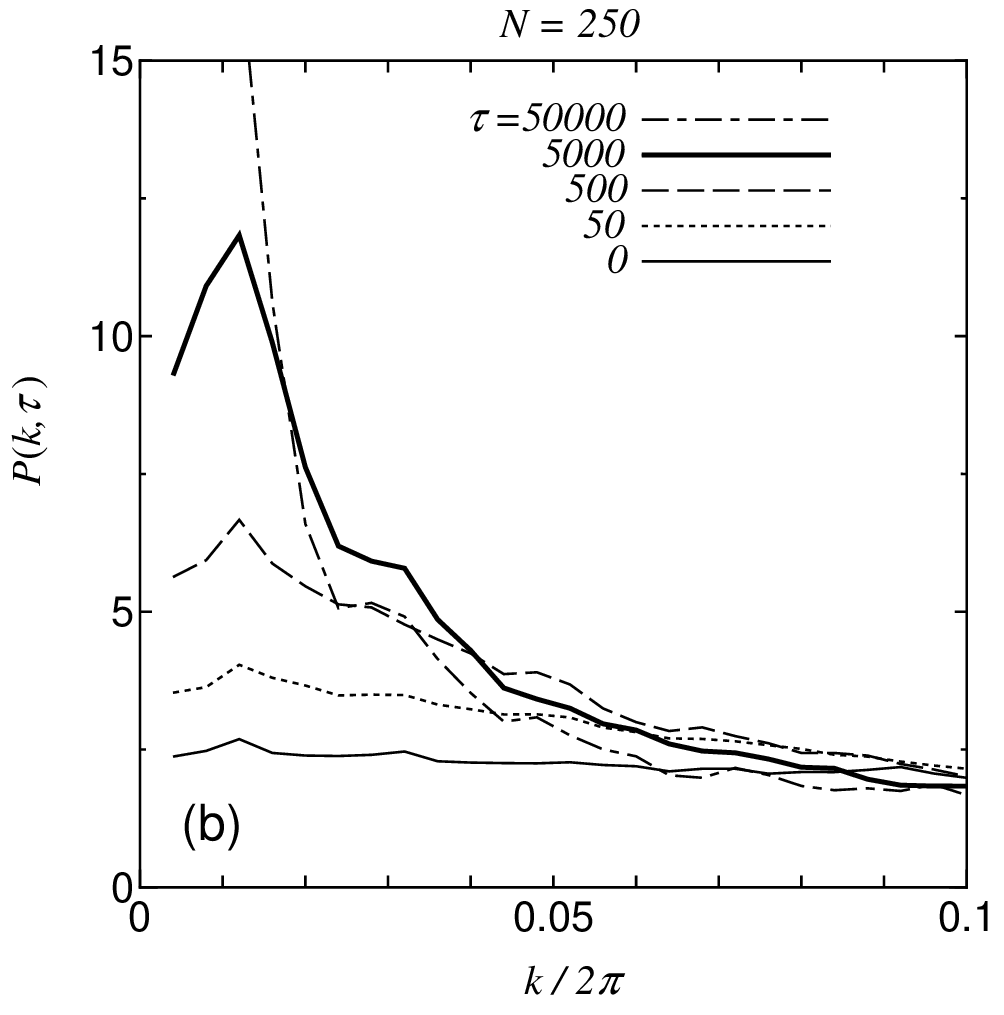}}
\caption{\protect
The intra-chain 
correlation function $E(n,\tau)$ 
of the non-bonded interactions  
defined by eq.25 in  (a),  
and its Fourier transformation   $P(k,\tau)$ defined by 
eq.26  in (b), for  $N=250$. 
Here $\tau=0$ (thin-solid line), 
 $5\times10$ (dotted line), $5\times10^2$ (dashed line), 
 $5\times10^3$ (bold line), and $5\times10^4$ (dot-dash line).
With increasing $\tau$ we can see development 
of the minimum and the maxima presumably due 
to entanglements. The inset in (a)  shows  $E(n,\tau)$ 
for small $n$ for these $\tau$. 
}
\label{fig6}
\end{figure}

\newpage
\widetext
\begin{figure}[t]
\centerline{\epsfxsize=5.in\epsfbox{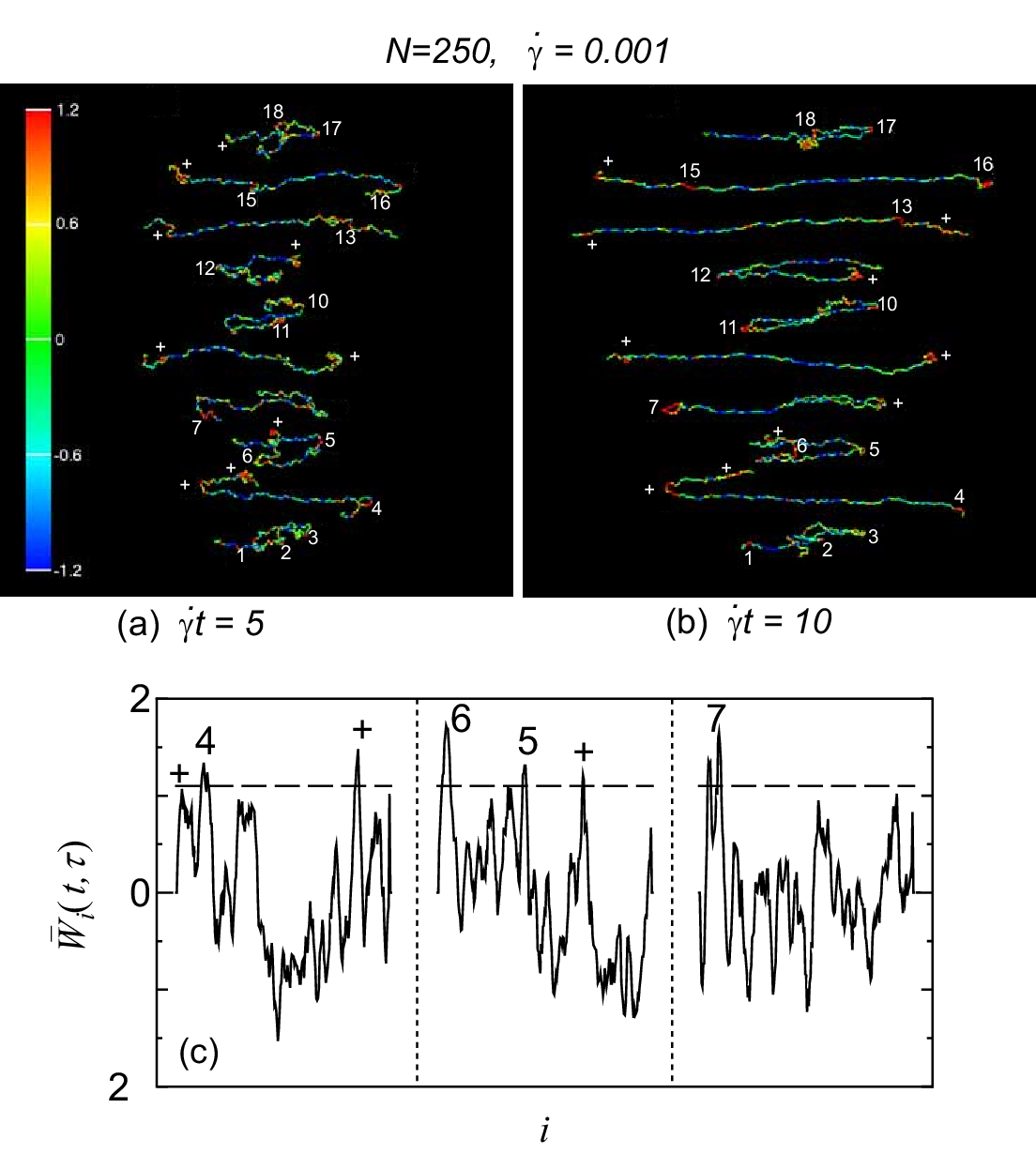}}
\caption{\protect
Snapshots of the deformed chains with $N=250$ 
on the $xz$ plane 
at $\gdot t = 5$ in (a) and 
at $\gdot t = 10$ in (b), where a 
 shear flow with  $\gdot=10^{-3}$ 
 was  applied  at $t=0$ with 
the same initial chain configuration as in Fig.5b. 
The non-bonded interactions 
with $\tau=500$ are written on the chains. 
The active spots satisfying the criterion 
given around  eq.24 are  detected, among which 
 the numbered segments correspond to those in Fig.5b. 
However, the active spots  which do not correspond to those 
in Fig.5b are marked by $+$.  
The flow is in the horizontal ($x$) direction, 
and the shear gradient  is 
 in the out-of-plane ($y$) direction. 
In (c) the data of 
$[W_i(t,\tau)- \av{W}] /\sigma (\tau)$  
for the three chains with the spots 4-7 in (b) are shown. 
}
\label{fig7}
\end{figure}

\newpage
\widetext
\begin{figure}[t]
\centerline{\epsfxsize=5.in\epsfbox{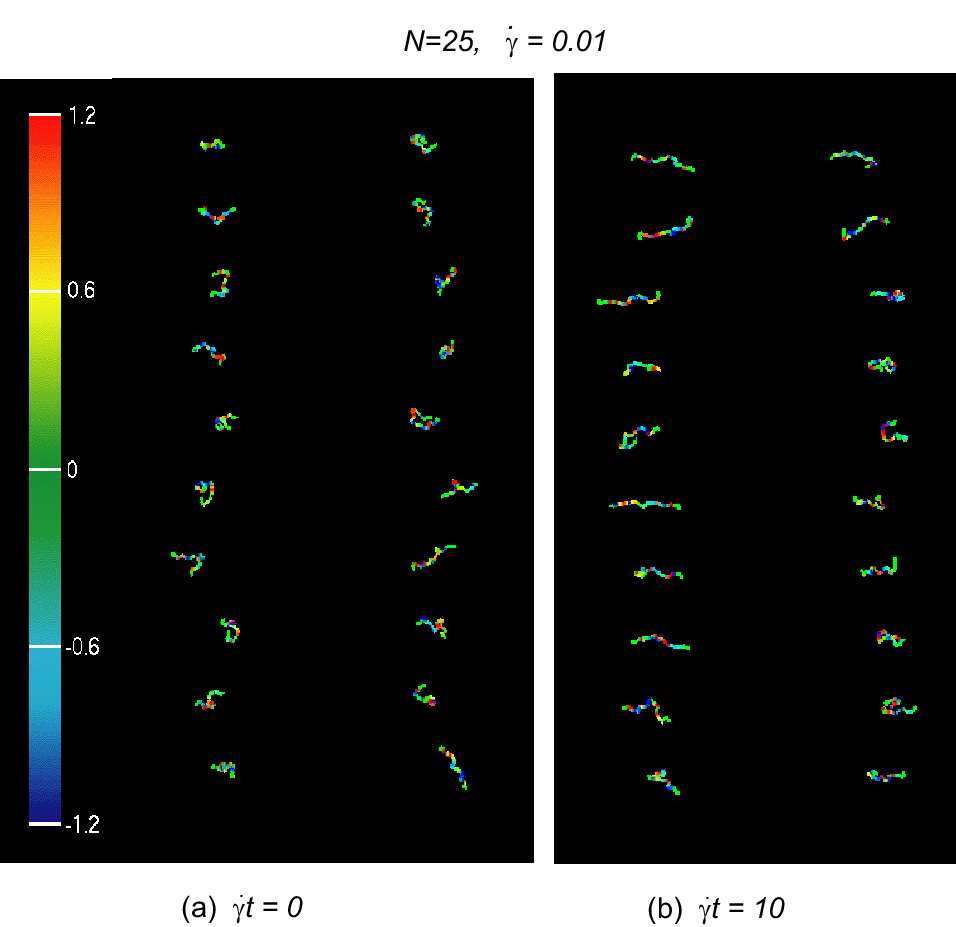}}
\caption{\protect
Snapshots of short chains 
with $N=25$ in the $xz$ plane 
at $\gdot t = 5$ in (a) and 
at $\gdot t = 10$ in (b), where a 
shear flow with  $\gdot=10^{-2}$ 
was  applied  at $t=0$. 
}
\label{fig8}
\end{figure}

\newpage
\narrowtext
\begin{figure}[t]
\centerline{\epsfxsize=2.6in\epsfbox{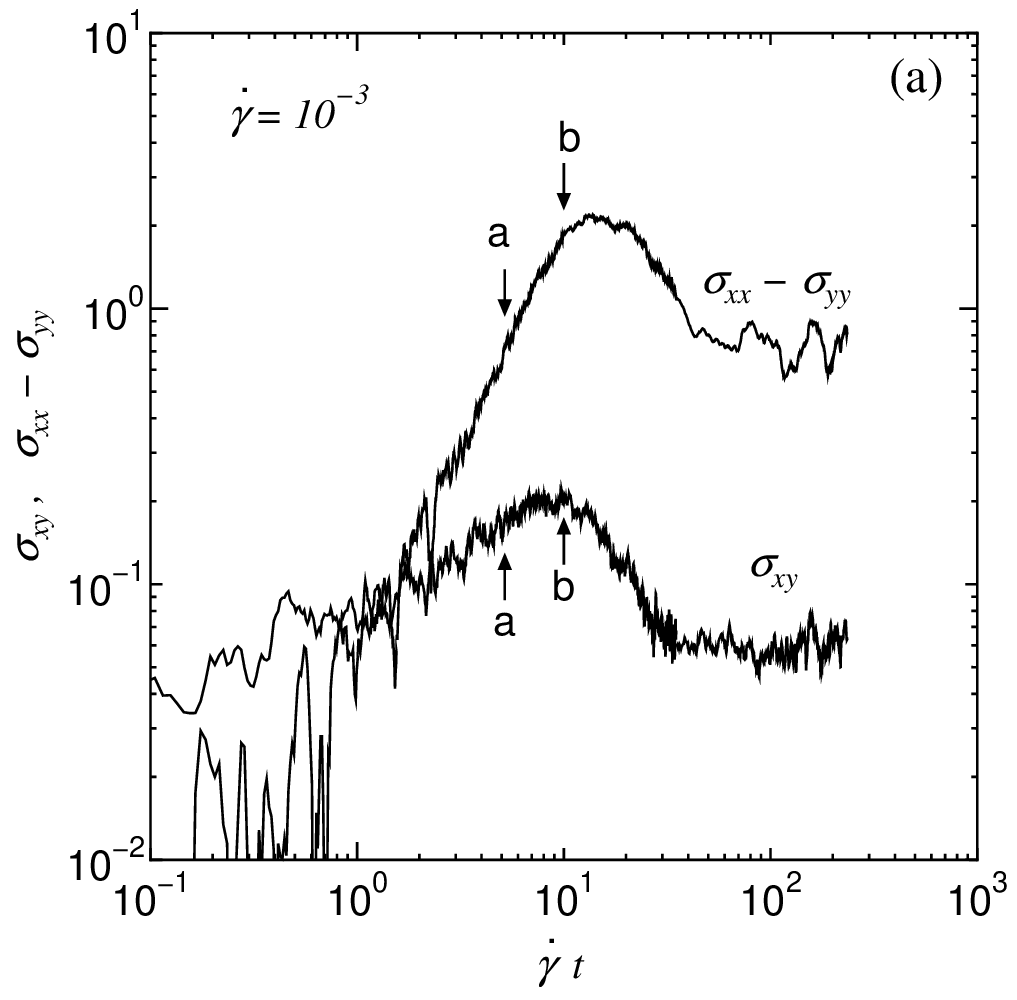}
\epsfxsize=2.6in\epsfbox{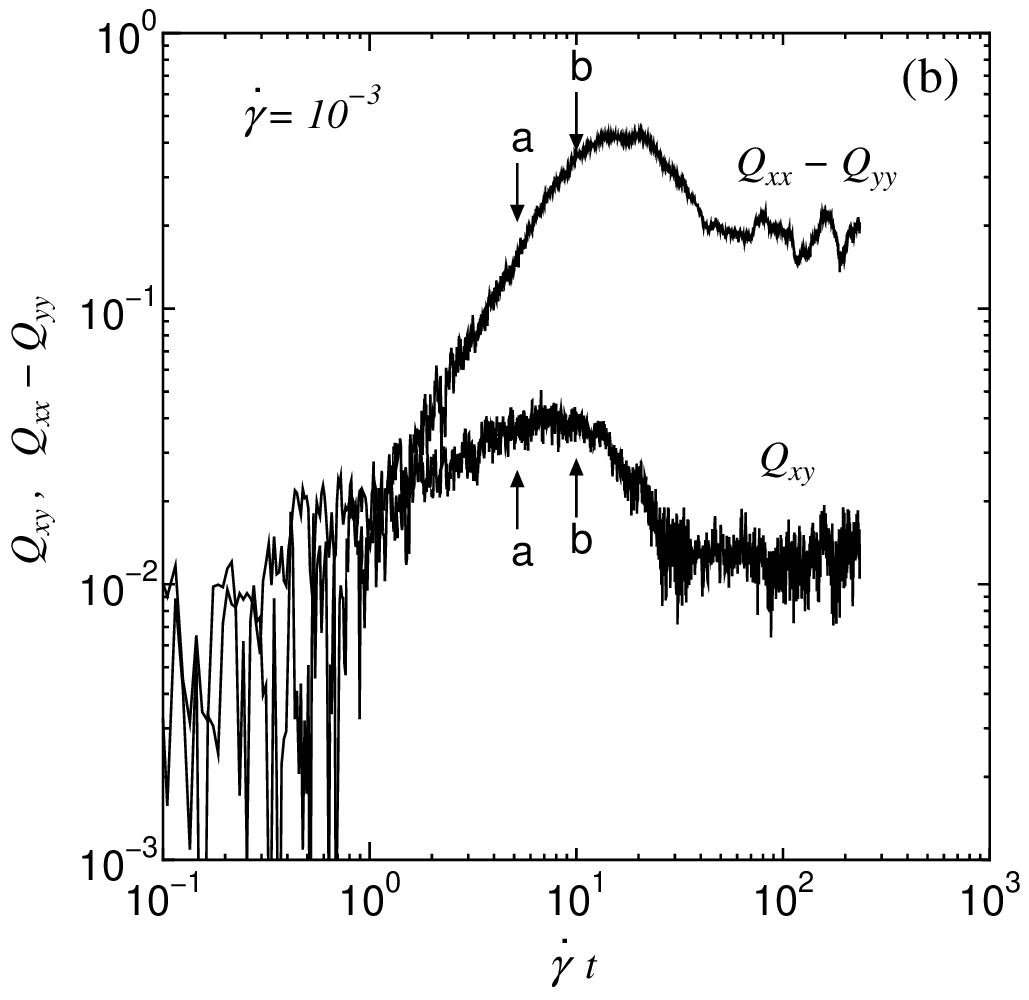}}
\centerline{\epsfxsize=2.6in\epsfbox{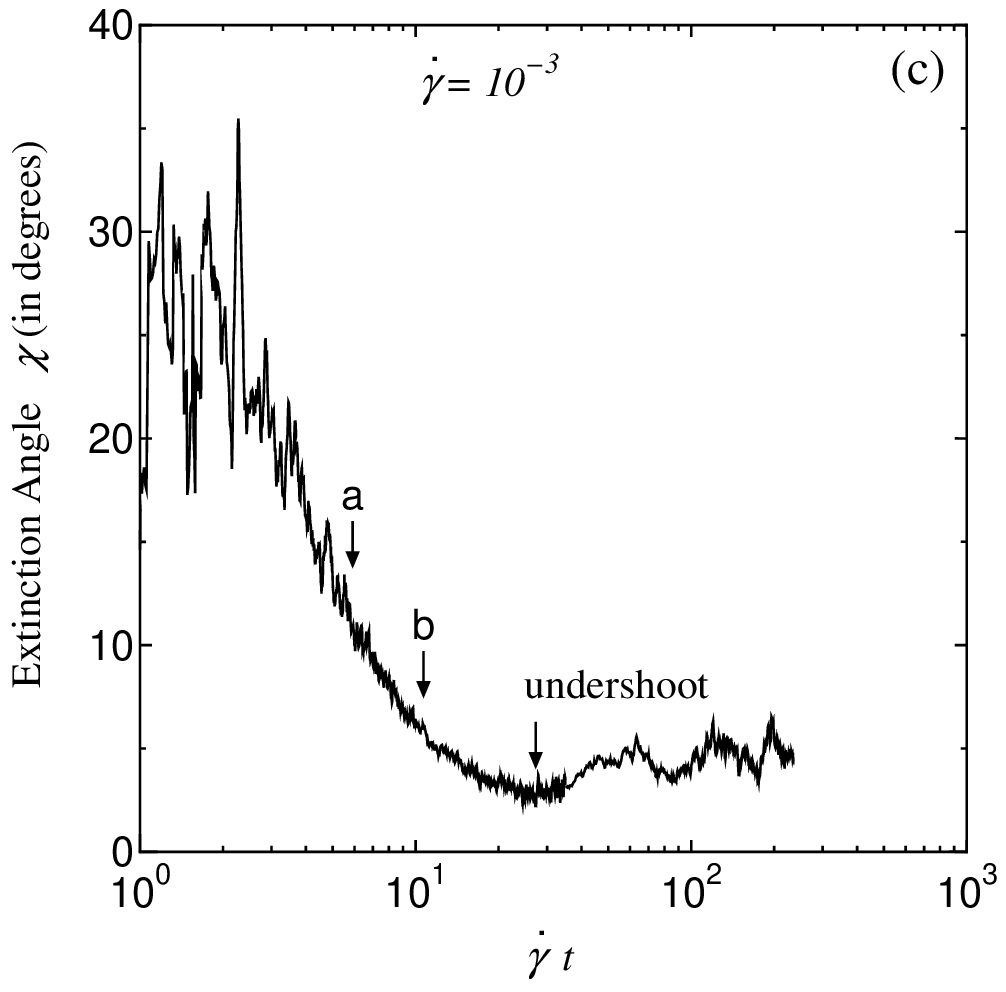}}
\caption{\protect
The stress growth functions after application of shear flow 
with   $\gdot=10^{-3}$ for $N=250$ in (a). The arrows 
(a and b) indicate the 
points of $\gdot t = 5$ in Fig.7a and 
 $\gdot t = 10$ in  Fig.7b.  The maximum of 
 $\sigma_{xy}(t)$ is at $\gdot t = 10$ and that of $N_1(t)$ 
 is at $\gdot t = 20$. 
   The corresponding components 
 of the orientational tensor $Q_{\alpha \beta}(t)$ in eq.27 
  are shown in (b). 
The extinction angle $\chi$ in eq.30 
is  shown in (c), which undershoots before reaching 
a larger steady-state value. 
These figures are the results of the single run 
displayed in Fig.7.  
}
\label{fig9}
\end{figure}

\newpage
\widetext
\begin{figure}[t]
\centerline{\epsfxsize=5.in\epsfbox{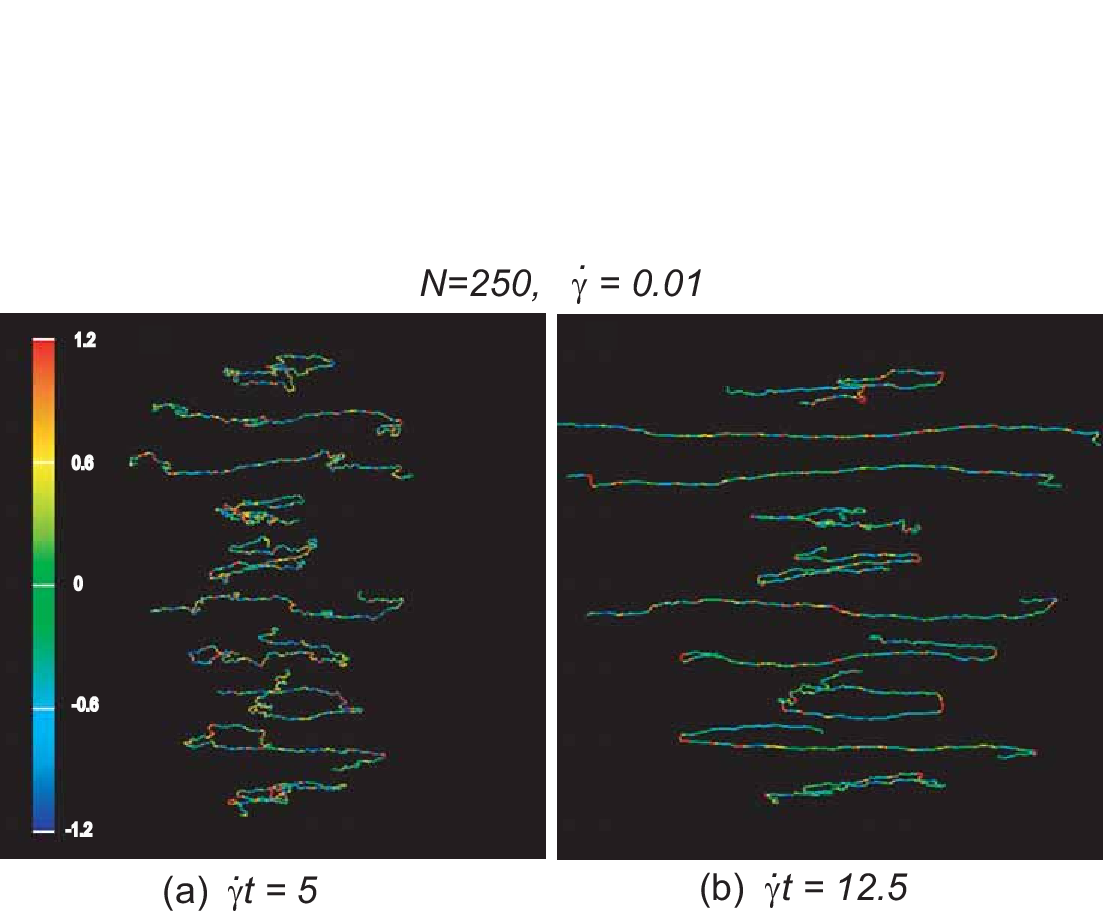}}
\caption{\protect
Snapshots of the deformed chains with $N=250$ 
in the $xz$ plane 
at $\gdot t = 5$ in (a) and 
at $\gdot t = 12.5$ in (b), where a 
 shear flow with  $\gdot=10^{-2}$ 
 was  applied  at $t=0$. 
The non-bonded interactions 
with $\tau=50$ are written on the chains using the color map 
on the left.  
In (b) the stretching is nearly complete 
between the bends. 
}
\label{fig10}
\end{figure}

\newpage
\widetext
\begin{figure}[t]
\centerline{\epsfxsize=2.8in\epsfbox{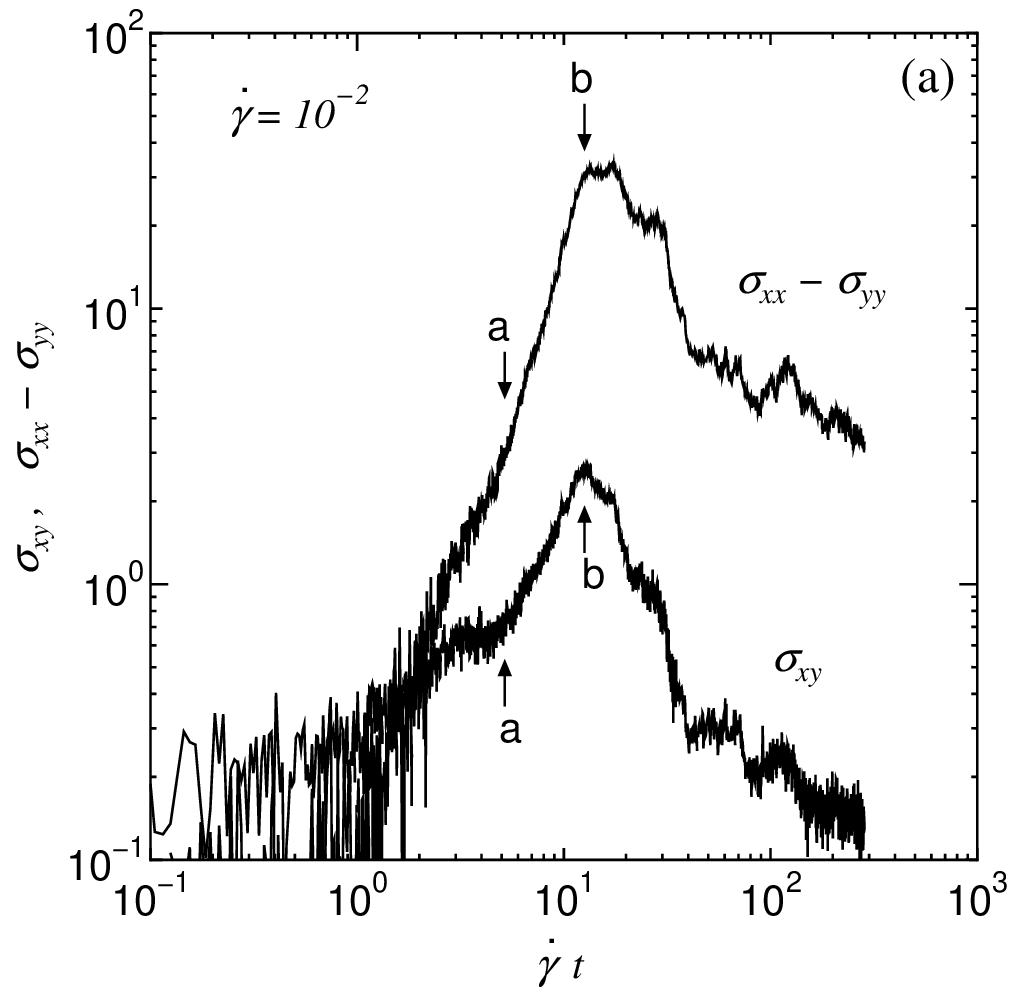}\epsfxsize=2.8in\epsfbox{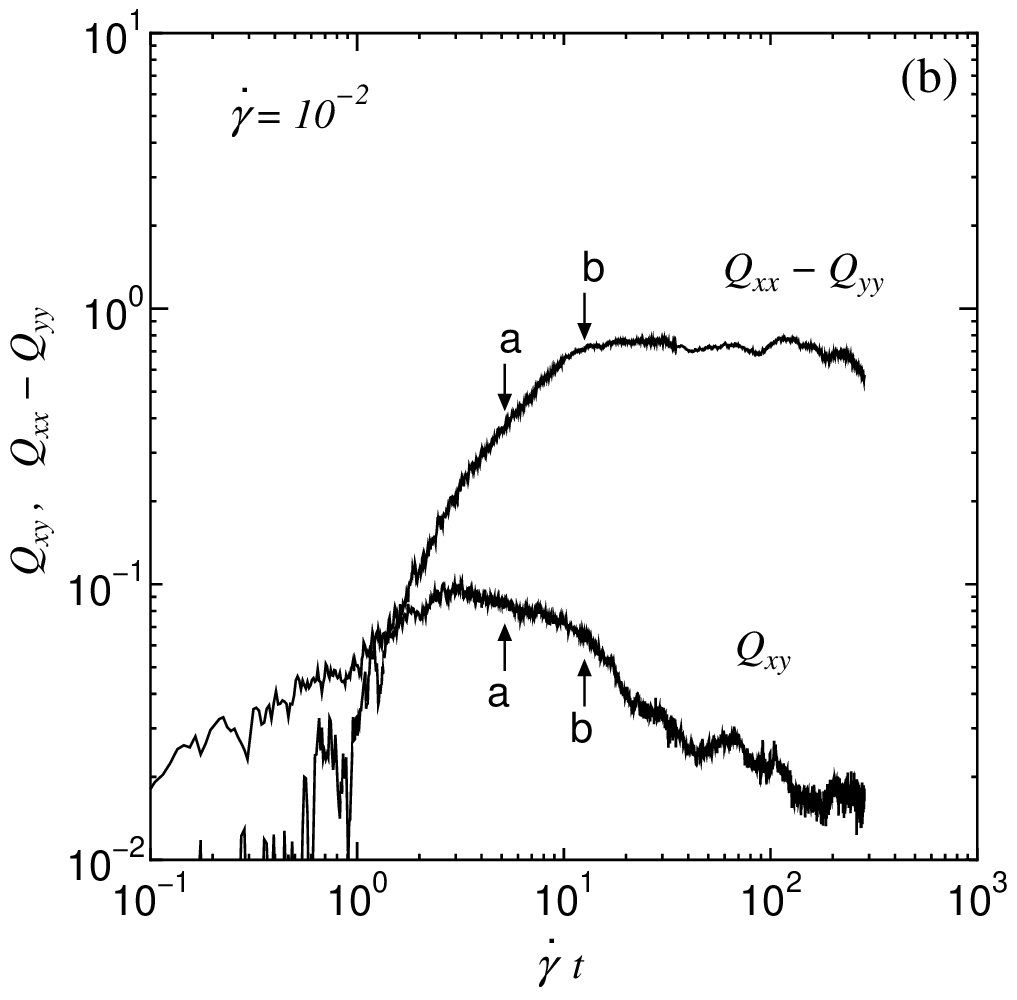}}
\caption{\protect
The stress growth functions after application of shear flow 
with   $\gdot=10^{-2}$ for $N=250$ in (a). The arrows 
(a and b) indicate the 
points of $\gdot t = 5$ in Fig.10a  and 
 $\gdot t = 12.5$ in  Fig.10b. 
  The corresponding components 
 of the orientational tensor $Q_{\alpha \beta}(t)$ are 
shown in (b), which demonstrate 
 bond elongation along the flow and 
behave very differently from the stress components. 
}
\label{fig11}
\end{figure}


\begin{references}
\bibitem{yo2} 
If a chain system is  near the glass transition,   
the stress  relaxation becomes extremely slow 
and nonlinear shear regimes 
are realized  even at very small shear rates.
See  calculations of $G(t)$  and rheology 
in such a system \cite{yo1,Onukibook}.
\bibitem{comment4} 
For example, the fluctuation of $\av{|{\bi P}(t_0)|^2}$ 
as a function of $t_0$  was  of order $6\%$ 
for $N=10$ and $M=100$ and 
was of order $25 \%$ for $N=250$ and $M=10$, 
as compared to the time averages over a time interval of 
$10^6$.
\bibitem{comment} If $b$ is determined by eq.4, 
it   becomes very weakly dependent on $N$ as 
$b=1.04 (N-1)^{0.05}$ in our simulations.
\bibitem{comment6} 
There is large ambiguity in the prefactor 
of $\tau_{\rm e}$ \cite{schleger}. 
That is,   $\tau_{\rm e}= 
(\pi/12)\zeta b^2 N_{\rm e}^2 /k_{\rm B}T
$ if $\ell(\tau_{\rm e})= a= N_{\rm e}^{1/2}b$ \cite{doi}. 
However, $\tau_{\rm e}= 
(1/3\pi^2)\zeta b^2 N_{\rm e}^2 /k_{\rm B}T$ 
if $\tau_{\rm e}= \tau_{\rm R}$ for $N=N_{\rm e}$.
\bibitem{comment5} 
In  Ref. \cite{yo2}, we used   the same 
chain model with  $N=10$ and $n\sigma^3=1$ 
to  obtain $A_0= 2.2k_{\rm B}Tn$
for various $T$   as the proportionality coefficient 
of the relation in eq.28. See  Fig.6 of Ref. \cite{yo2}. 
This expression also holds in the present 
work and is thus  independent of $N$.
\end{references}
\end{document}